\def\aa{{A\&A}}

\def\actaa{{AcA}}

\def\apj{{ApJ}}
\def\esomess{{The Messenger}}

\def\newa{{NewA}}

\def\cm2{cm$^{-2}$}

\def\c2{C~{\sc ii}}
\def\c4{C~{\sc iv}}
\def\fe2{Fe~{\sc ii}}
\def\fe3{Fe~{\sc iii}}
\def\mg1{Mg~{\sc i}}
\def\mg2{Mg~{\sc ii}}
\def\si2{Si~{\sc ii}}
\def\si4{Si~{\sc iv}}
\def\al2{Al~{\sc ii}}
\def\al3{Al~{\sc iii}}
\def\o1{O~{\sc i}}
\def\n1{N~{\sc i}}
\def\h1{H~{\sc i}}

\def\approxlt{\mathrel{\spose{\lower 3pt\hbox{$\sim$}}
        \raise 2.0pt\hbox{$<$}}}
\def\approxgt{\mathrel{\spose{\lower 3pt\hbox{$\sim$}}
        \raise 2.0pt\hbox{$>$}}}

\documentclass[article]{gwp80}
\usepackage{graphicx}
\usepackage{amssymb}
\tabletypesize{\normalsize}  

\shortauthors{Catelan et al.}
\shorttitle{The VVV Survey}

\begin{document}
\large    
\pagenumbering{arabic}
\setcounter{page}{101}

\title{The Vista Variables in the V\'{i}a L\'actea (VVV) ESO Public Survey: 
Current Status \\ and First Results}

%
%
\author{{\noindent 
M. Catelan{$^{\rm 1}$}, 
D. Minniti{$^{\rm 1}$}, 
P. W. Lucas{$^{\rm 2}$}, 
J. Alonso-Garc\'{i}a{$^{\rm 1}$}, 
R. Angeloni{$^{\rm 1}$}, 
J. C. Beam\'{i}n{$^{\rm 1}$}, 
C. Bonatto{$^{\rm 3}$}, 
J. Borissova{$^{\rm 4}$}, 
C. Contreras{$^{\rm 1,2}$}, 
N. Cross{$^{\rm 5}$}, 
I. D\'ek\'any{$^{\rm 1}$}, 
J. P. Emerson{$^{\rm 6}$}, 
S. Eyheramendy{$^{\rm 1}$}, 
D. Geisler{$^{\rm 7}$}, 
E. Gonz\'alez-Solares{$^{\rm 8}$}, 
K. G. Helminiak{$^{\rm 1}$}, 
M. Hempel{$^{\rm 1}$}, 
M. J. Irwin{$^{\rm 8}$}, 
V. D. Ivanov{$^{\rm 9}$},
A. Jord\'an{$^{\rm 1}$}, 
E. Kerins{$^{\rm 10}$}, 
R. Kurtev{$^{\rm 4}$},
F. Mauro{$^{\rm 7}$}, 
C. Moni Bidin{$^{\rm 7}$}, 
C. Navarrete{$^{\rm 1}$}, 
P. P\'erez{$^{\rm 1,11}$}, 
K. Pichara{$^{\rm 1}$}, 
M. Read{$^{\rm 5}$}, 
M. Rejkuba{$^{\rm 12}$}, 
R. K. Saito{$^{\rm 1}$}, 
S. E. Sale{$^{\rm 1,4}$}, 
I. Toledo{$^{\rm 1}$} 
\\
\\
{\it 
(1) Pontificia Universidad Cat\'olica de Chile, Santiago, Chile\\
(2) University of Hertfordshire, Hatfield, UK \\
(3) Universidade Federal do Rio Grande do Sul, Porto Alegre, Brazil\\
(4) Universidad de Valpara\'{i}so, Valpara\'{i}so, Chile \\
(5) The University of Edinburgh, Edinburgh, UK \\
(6) Queen Mary, University of London, London, UK \\
(7) Universidad de Concepci\'on, Concepci\'on, Chile \\
(8) University of Cambridge, Cambridge, UK \\
(9) European Southern Observatory, Santiago, Chile \\
(10) The University of Manchester, Manchester, UK \\ 
(11) Universidad de Chile, Santiago, Chile \\
(12) European Southern Observatory, Garching, Germany \\
}
}
}

%
%
\email{(1) mcatelan,dante@astro.puc.cl (2) p.w.lucas@herts.ac.uk}


\begin{abstract}
Vista Variables in the V\'{i}a L\'actea (VVV) is an ESO Public Survey that is performing a variability survey of the Galactic bulge and part of the inner disk using ESO's Visible and Infrared Survey Telescope for Astronomy (VISTA). The survey covers $520 \, {\rm deg}^2$ of sky area in the $ZYJHK_S$ filters, for a total observing time of 1929 hours, including $\sim 10^9$ point sources and an estimated $\sim 10^6$ variable stars. Here we describe the current status of the VVV Survey, in addition to a variety of new results based on VVV data, including light curves for variable stars, newly discovered globular clusters, open clusters, and associations. A set of reddening-free indices based on the $ZYJHK_S$ system is also introduced. Finally, we provide an overview of the VVV Templates Project, whose main goal is to derive well-defined light curve templates in the near-IR, for the automated classification of VVV light curves.  
\end{abstract}

\section{Introduction}\label{sec:intro}
The Vista Variables in the V\'{i}a L\'actea (VVV) ESO Public Survey\footnote{\tt{http://vvvsurvey.org/}} is an infrared variability survey of the Milky Way bulge and an adjacent section of the mid-plane where star formation activity is high \citep{dmea10,rsea10}. The survey is carried out on ESO's Visible and Infrared Survey Telescope for Astronomy (VISTA) 4m telescope, which is equipped with a wide-field near-infrared (IR) camera (VIRCAM; Dalton et al. 2006) that is a mosaic of 16 $2048 \times 2048$ detectors, each sensitive over the spectral region $\lambda = 0.8 - 2.5 \, \mu {\rm m}$, and with an average pixel scale of $0.34 \, '' {\rm pixel}^{-1}$. The total (effective) field of view (FoV) of the camera is $1.1 \times 1.5 \, {\rm deg}^2$. 

\begin{figure*}
\centering
\includegraphics[width=\textwidth]{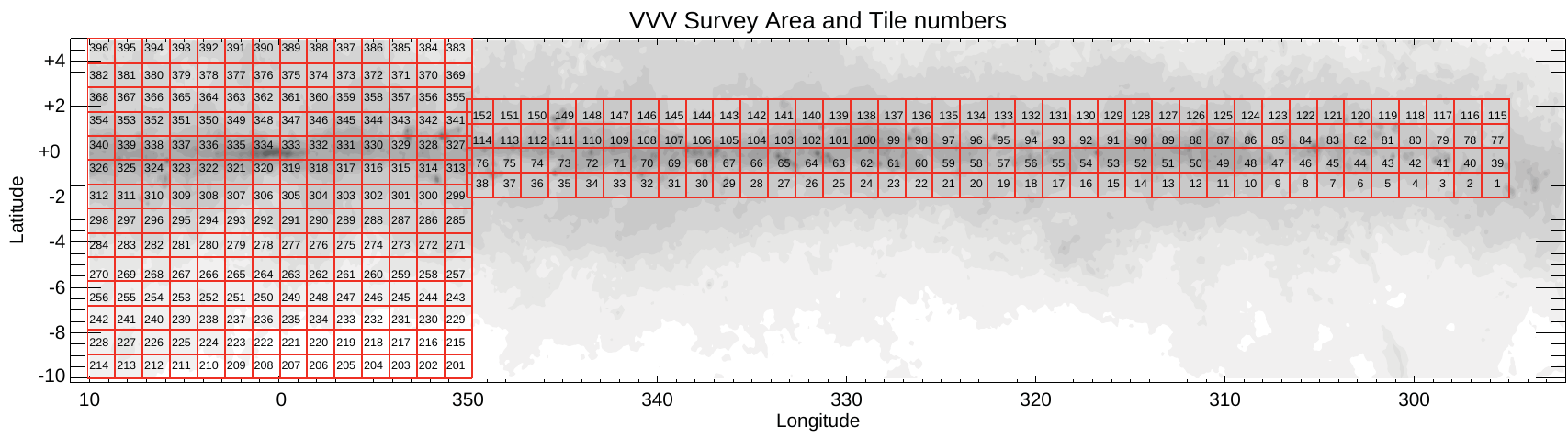}
\vskip0pt
\caption{The VVV Survey area, with individual ``tiles'' (i.e., individual observation fields, obtained from a combination of ``pawprints''; see, e.g., \citeauthor{dmea10} \citeyear{dmea10}) numbered. This is a plot of Galactic latitude $b$ version Galactic longitude $\ell$, overplotted on a differential extinction contour map.}
\label{fig:vvvfield}
\end{figure*}

In addition to providing images covering a wide range in effective wavelength ($ZYJHK_S$ filters), the VVV Survey is able to probe much deeper into high extinction regions than its predecessors. In particular, near-IR color-magnitude diagrams (CMDs) of some VVV fields extend about 4 magnitudes deeper than their 2MASS counterparts \citep{rsea10}, and also about 1 magnitude deeper than the UKIDSS Galactic Plane Survey (GPS) in the regions of overlap \citep{dmea10}. Most importantly, the VVV Survey will also provide a window into time-variable phenomena, by repeatedly covering the Galactic bulge and regions of the inner plane over a timespan of about 5 years, with a total of 1929 hours of observing time. Of order $10^9$ point sources within a total sky area of $520 \, {\rm deg}^2$ will thus be monitored. According to the December 2010 version of the \citet{wh96} catalog, the surveyed region includes 36 globular clusters\footnote{An additional globular cluster, FSR1767 \citep{cbea07}, is also in the surveyed region, but does not yet appear in the \citet{wh96} catalog (see also \S\ref{sec:globulars}).} and $\sim 314$ known open clusters, and is shown in Figure~\ref{fig:vvvfield}. The final products will be a deep IR atlas in 5 passbands and a catalogue containing an estimated $\sim 10^6$ variable stars. The synergy with other surveys that provide optical, mid-IR, and far-IR data is also worth noting, including UKIDSS GPS, IPHAS, VPHAS+, the Spitzer, GLIMPSE, and MIPSGAL surveys, and the all-sky AKARI and WISE surveys. 

The scientific goals of the VVV Survey are described in great detail in \citet{dmea10}. One of the main goals is to gain insight into the inner Milky Way's origin, structure, and evolution. This will be achieved by, for instance, obtaining a first, precise 3D map of the Galactic bulge, whose detailed shape appears to be quite complex \citep[e.g.,][]{mcwz10,dnea10,rdpea11}. Of particular importance, in order to achieve this goal, will be the RR Lyrae stars. Not only are the RR Lyrae variables present in the direction of the bulge in great numbers \citep{isea11}, but they also follow precise period-luminosity relations in the near-IR bandpasses \citep[e.g.,][]{mcea04}. Moreover, being very old, they are ``living eyewitnesses'' of virtually the entire formation history of the Milky Way \citep[see, e.g.,][for a recent review]{mc09}. Another important goal of the VVV Survey is the discovery of globular clusters, open clusters, and stellar associations which have heretofore remained ``hidden'' from view behind the high extinction regions towards the inner Galaxy. In this paper, we provide an update on the current status of the project, as well as a first look into some of the science that has so far been enabled by the VVV data. 

\begin{flushleft}
\begin{deluxetable*}{rccc}
\tabletypesize{\normalsize}
\tablecaption{Overview of VVV Observations: \\First Year}
\tablewidth{0pt}
\tablehead{ \\ \colhead{Filters} & \colhead{Planned\tablenotemark{a}}   & \colhead{Executed\tablenotemark{a}} &  \colhead{Percentage} \\
}
\startdata
\multicolumn{4}{c}{Bulge} \\
\hline\\
$JHK_S$               & $196$ & $176$ & $90\%$ \\ 
$ ZY$                 & $196$ & $ 78$ & $40\%$ \\ 
var\tablenotemark{b}  & $980$ & $113$ & $12\%$ \\
\\
\hline\\
\multicolumn{4}{c}{Disk} \\
\hline\\
$JHK_S$               & $152$ & $152$ & $100\%$ \\ 
$ ZY$                 & $152$ & $128$ & $ 84\%$ \\ 
var\tablenotemark{b}  & $760$ & $547$ & $ 72\%$ \\
\enddata
\tablenotetext{a}{In units of observing blocks (OBs).}
\tablenotetext{b}{For the first year, the plan was to execute a total of 5 observations for each of the 196 (152) bulge (disk) fields, giving a total of 980 (760) OBs.}
\label{tab:completion}
\end{deluxetable*}
\end{flushleft}

\section{The VVV Survey: Current Status}\label{sec:status}

\subsection{Pipeline Processing at CASU}\label{sec:pipeline}
VVV observations are carried out in service mode. During the first year, $JHK_S$ images typically had $0.8''$ seeing, whereas $ZY$ images typically had a seeing of $1''$. Once observations are carried out, the VVV data are processed by a pipeline at the Cambridge Astronomy Survey Unit \citep[CASU; see, e.g.,][]{miea04,jlea10}. In addition to properly reduced and calibrated images, the CASU pipeline also delivers aperture photometry for the observed fields. At the same time, members of our team perform profile-fitting photometry for the denser fields (e.g., star clusters). Work is currently in progress also to enable difference-imaging analyses \citep[e.g.,][]{ca00,db08} in the multi-epoch, $K_S$-band data, for the detection of stellar variability in selected fields.   

CASU has recently implemented v1.1 of its pipeline, and completed processing all the data obtained so far with the VISTA telescope. Some of the improvements that went into the new version of the pipeline include the following:\footnote{See also {\tt  http://casu.ast.cam.ac.uk/surveys-projects/vista/data-processing/version-log} for a comparison between different versions of the CASU pipeline.} 

\begin{itemize}
\item Improved calculation of aperture corrections in very crowded fields, which goes into the computation of the photometric zero points.

\item Calculation of separate zero points for each of the $6\times 16=96$ sections of a VISTA tile. This corrects for both the significant non-uniformity of the VIRCAM point-spread function (PSF) across the FoV and changes in the PSF between the 6 pawprints that make up a tile. (See \S\ref{sec:varsvf} for a description of ``pawprints'' and ``tiles.'')

\item Improved correction for the effect of extinction on the photometric transformations from 2MASS to VISTA, which is part of the zero point calculation.

\item First data release of FITS images, confidence maps, and FITS single-band source catalogues will be released to ESO by the end of April/early May.
\end{itemize}

\subsection{Current Status of Observations}\label{sec:observations}
In Table~\ref{tab:completion}, we show the completion rate for the planned VVV first year observations, as of October 2010. Clearly, while almost complete coverage of the planned fields in $JHK_S$ has been obtained, allowing the production of deep near-IR CMDs, the same is not true for the $ZY$ filters, for which several fields could not be observed during the first year. Note also that the bulk of the planned variability campaign could not be carried out during the first year, especially in the case of the bulge fields. 

Unfortunately, the coating of the VISTA telescope mirrors has been degrading more rapidly than anticipated, leading to a loss in sensitivity, particularly at short wavelengths. As a consequence, operation of the VISTA telescope has been halted recently, so that the mirror coatings can be replaced. This means that VVV observations for ESO Period 86 will not be completed as expected, likely leading to an extension of our original 5-year goal for completion of the project.  

This notwithstanding, the available data already represent a treasure trove that can be explored in a variety of ways. In what follows, we report on some of the first scientific results that have been obtained on the basis of the VVV data acquired so far.

\begin{figure}[t]
\centering
\includegraphics[width=12cm]{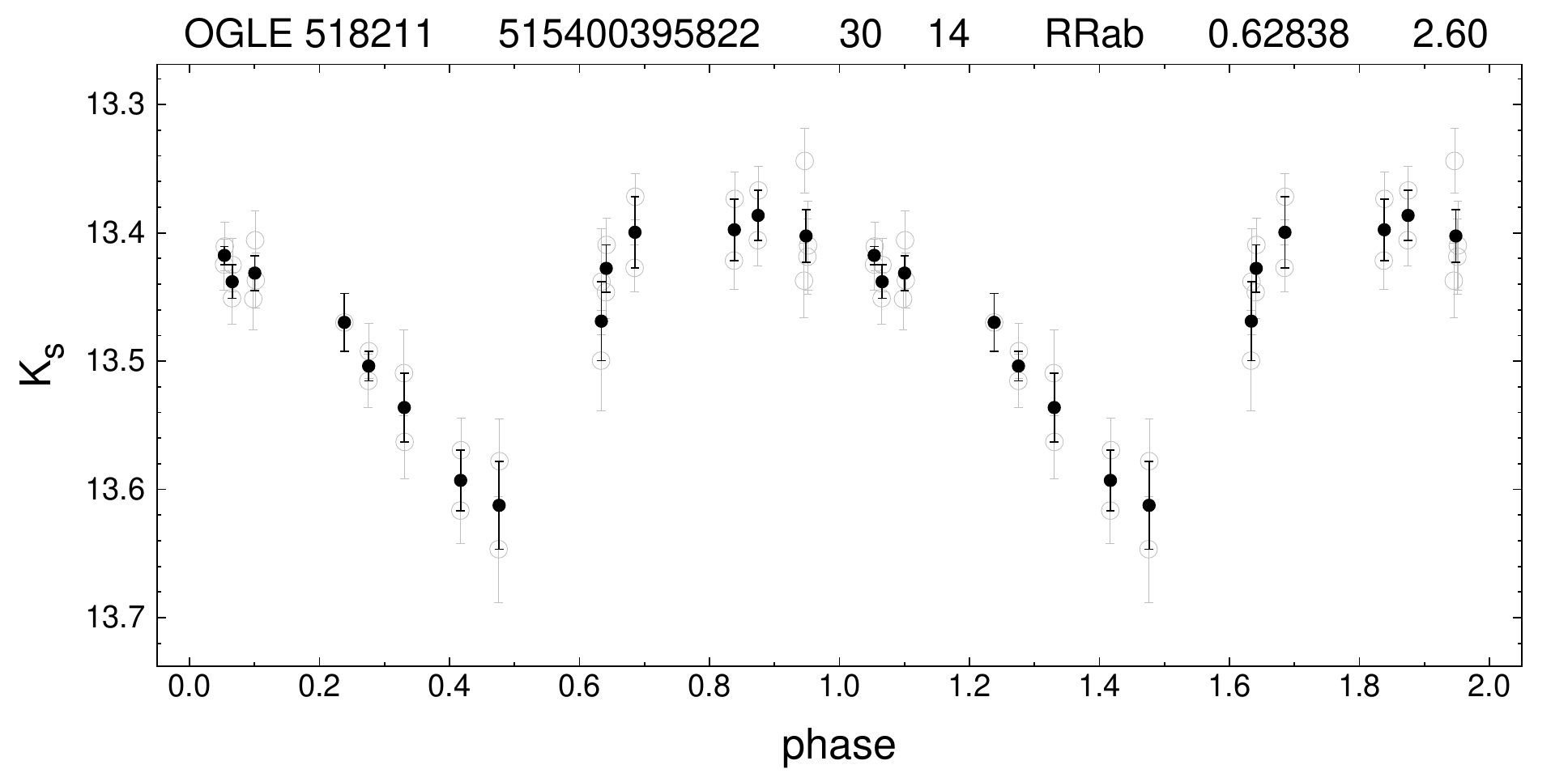}
\includegraphics[width=12cm]{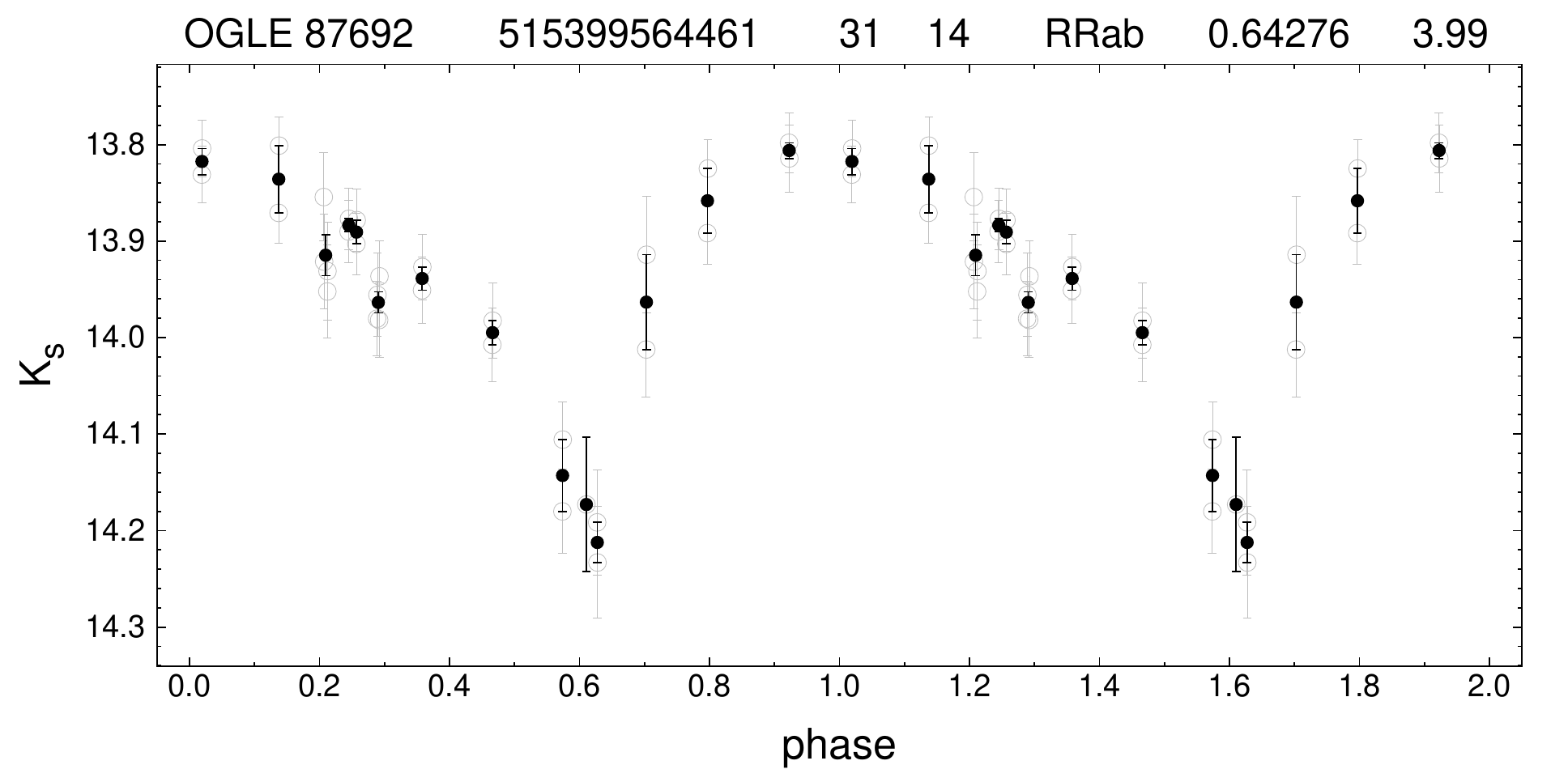}
\includegraphics[width=12cm]{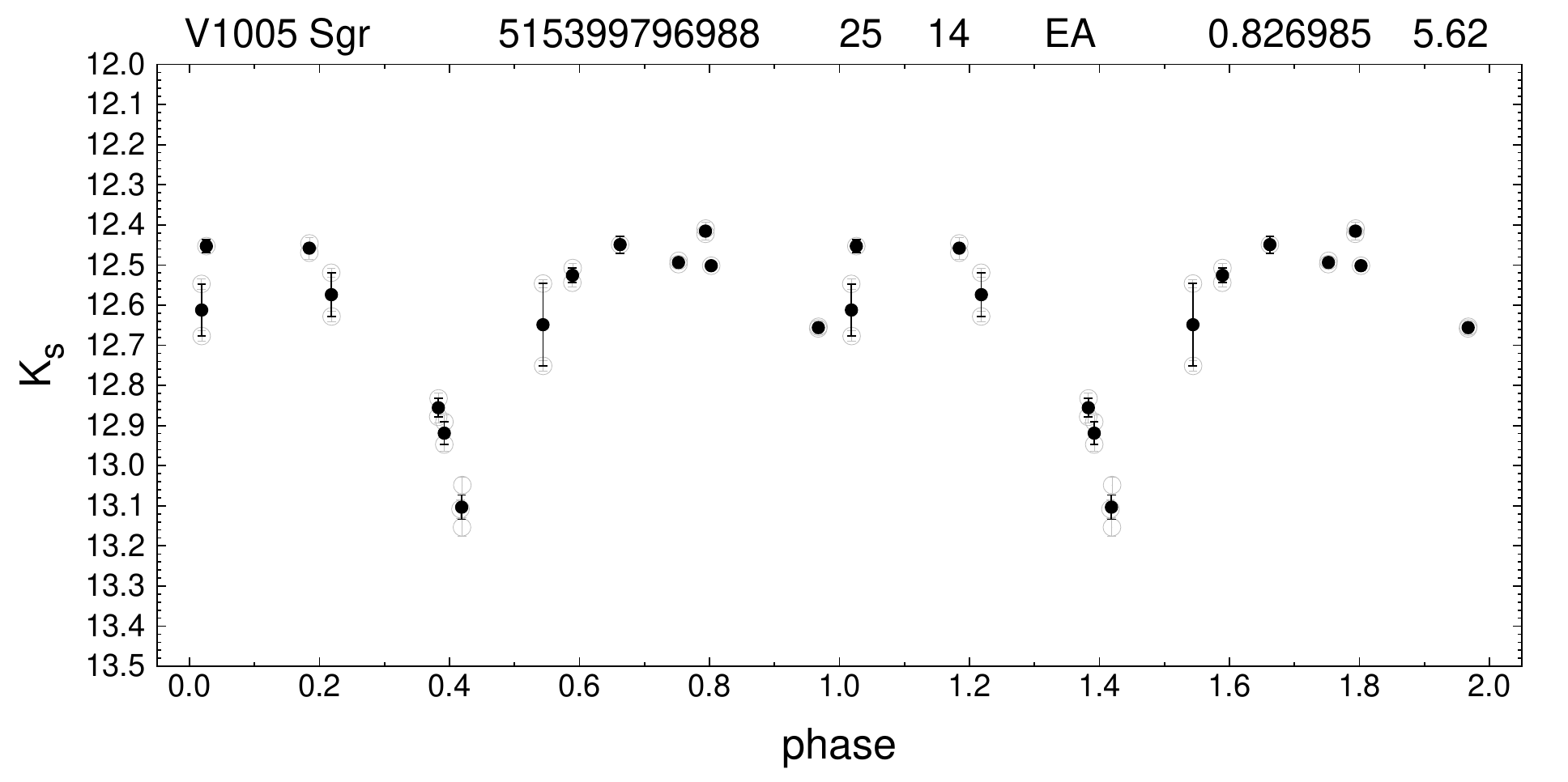}
\vskip0pt
\caption{Light curve for three variable stars in the VISTA science verification fields. {\em Top:} OGLE 518211, an ab-type RR Lyrae star. {\em Middle:} OGLE 87692, another ab-type RR Lyrae. {\em Bottom:} V1005 Sgr, a detached eclipsing binary. For all plots, from left to right, the following information is given: star ID in the OGLE catalog or in the GCVS; internal VVV source ID; number of data points in gray; number of data points in black; variability type; period in days; and a variability parameter $d$. See text for further details. }
\label{fig:svf}
\end{figure}

\section{The VVV Survey: First Results}\label{sec:results}

\subsection{Variable Stars in the Science Verification Fields}\label{sec:varsvf}

As can be seen from Table~\ref{tab:completion}, the completion rate of planned observations for our variability campaign is still quite low. However, as part of ESO's science verification of the VISTA telescope, additional data for our fields were obtained, in the period 19-30 Oct., 2009 \citep{maea10}. This consisted of multi-color imaging, in the $ZYJHK_S$ filters, of a field located at Galactic coordinates $b = -3.1^{\circ}$, $\ell = 2.2^{\circ}$, supplemented with an additional set of 13 $K_S$-band images of the same field. We have searched the images available for this field for the presence of known variable stars, from the MACHO and OGLE catalogs \citep{caea98,isea11} and the General Catalog of Variable Stars \citep[GCVS;][]{pkea98}, and are able to identify several previously catalogued variables, including RR Lyrae stars and eclipsing variables. In Figure~\ref{fig:svf}, we show three examples of the light curves obtained. This includes the ab-type RR Lyrae stars OGLE 518211 ({\em upper panel}) and OGLE 87692 ({\em middle panel}), additionally to the Algol-type (detached) eclipsing binary V1005 Sgr ({\em bottom panel}). 

In Figure~\ref{fig:svf}, gray symbols denote the individual measurements (and their formal errors) by standard VISTA Science Archive [VSA] aperture photometry, performed on single VISTA images called ``pawprints.'' The latter have a non-contiguous sky coverage due to the gaps between the 16 chips of VIRCAM, resembling the pawprint of an animal (for an example, see Fig.~\ref{fig:omegacen} below, right panel). Around each pointing of VISTA, 6 pawprints with proper offsets are acquired in succession, in order to provide a contiguous sampling of a sky area; stacks of these images are called ``tiles.'' Thus, a point source is observed at least two and up to six times in a row (depending on its position), within a short time interval (5\,min 8\,sec for the bulge in the first year). Magnitude values corresponding to such exposure sequences were averaged, and are shown as black dots in Figure~\ref{fig:svf}, with error bars representing the standard deviation of the mean.

The time between the acquisition of two consecutive pawprints that correspond to the same tile is much shorter than the timescale of the light variation for the most typical variable stars that will be detected in the course of the VVV Survey. Therefore, these successive observations carry information about the internal scatter of each individual photometric time-series. The ratio $d$ between the global rms scatter of a light curve and the internal scatter measured in the above way provides a good means for determining the probability of a star being variable within the photometric accuracy. Objects with $d$ values deviating substantially from unit are good variable star candidates (see headers of Fig.~\ref{fig:svf}).

\begin{figure}[t]
\centering
    \includegraphics[width=\textwidth]{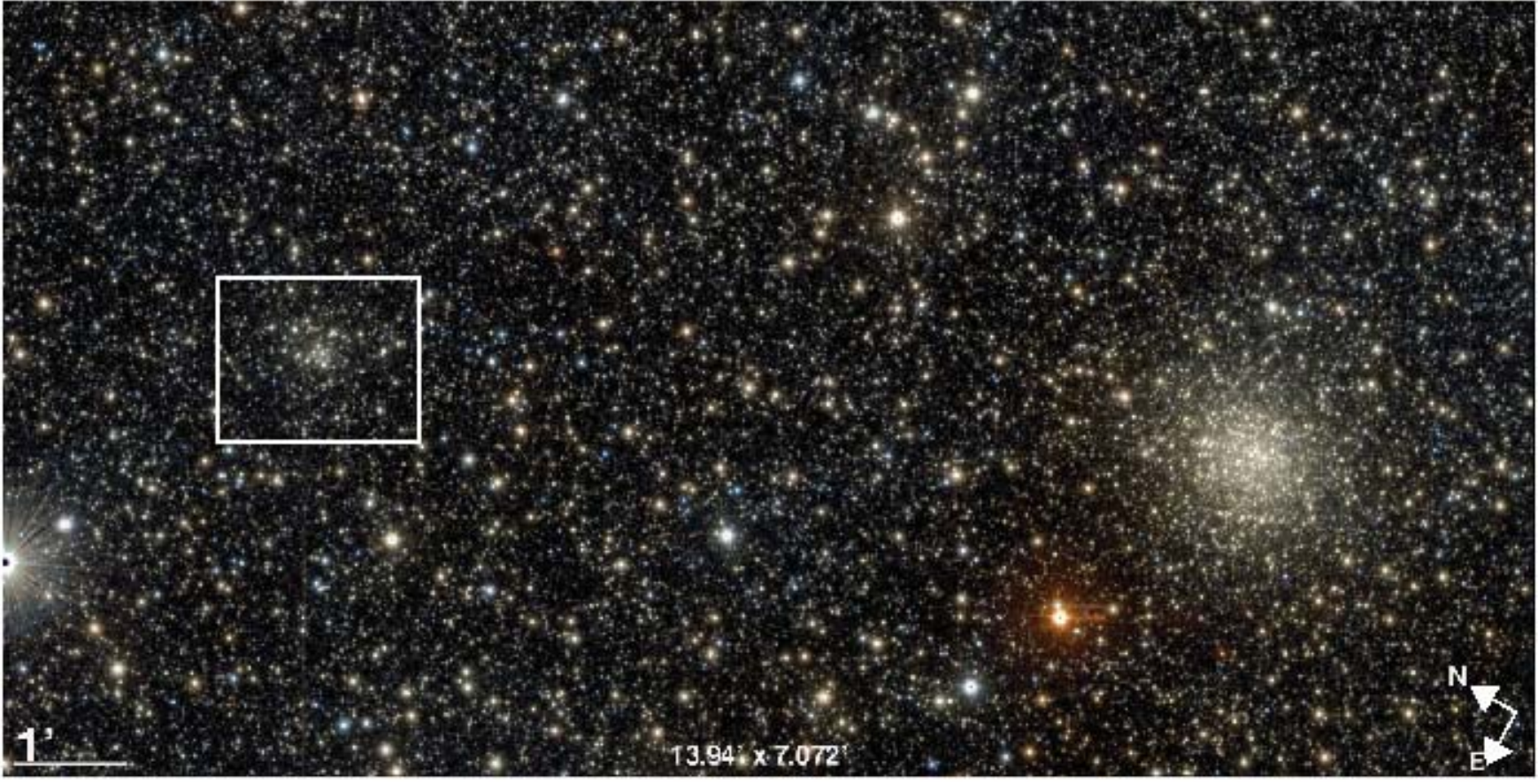}
\caption{$JHK_S$ color-composite image showing the newly discovered globular cluster VVV CL001 (framed in a white rectangle) in the vicinity of the previously known globular cluster UKS~1. From \citet{dmea11}. 
\label{fig:vvvcl001}}
\end{figure}

\subsection{Globular Star Clusters}\label{sec:globulars}

The census of Galactic globular clusters is still far from being complete, because distant objects can easily hide behind/amongst the very crowded and highly reddened stellar fields in the direction of the Galactic bulge and disk. Indeed, the asymmetry of the spatial distribution of known globular clusters around the Galactic center clearly indicates that the census of globular clusters towards the inner Milky Way is likely incomplete. In this sense, \citet{viea05} estimated that there should be $10\pm 3$ such clusters in the more central regions of the Galaxy. The advent of the new generation of extensive, wide-field surveys such as SDSS \citep{kaea09}, 2MASS \citep{msea06}, and GLIMPSE \citep{rbea03} permitted the detection of several new Galactic globular clusters. The December 2010 compilation of the \citet{wh96} catalog included seven new globulars not present in the February 2003 version, but six more objects have been proposed in the last years: SDSSJ1257+3419 \citep{sh06}, FSR584 \citep{ebea07}, FSR1767 \citep{cbea07}, FSR190 \citep{dfea08}, and Pfleiderer 2 \citep{soea09}.

\begin{figure}[t]
\centering
    \includegraphics[width=0.875\textwidth]{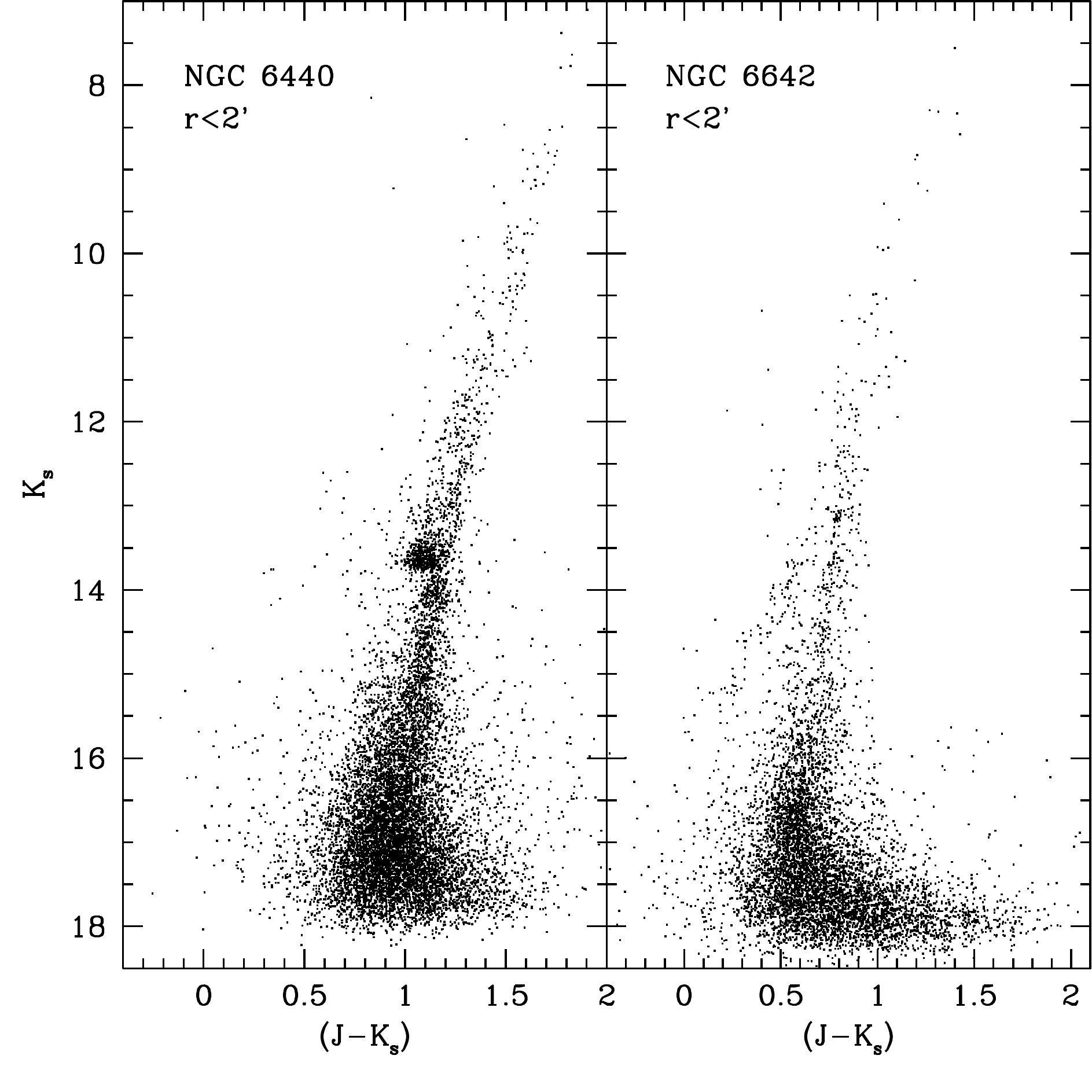}
\caption{CMDs of the inner $2'$ region of NGC~6440 and NGC~6626, two of the clusters sampled in the VVV Survey, representative of the metal-rich/moderately metal-poor globular cluster populations observed in the direction of the inner Galaxy.
\label{fig:6440-6642}}
\end{figure}

As far as detection of new globular cluster candidates goes, the VVV Survey should naturally play an important role. This has already been confirmed in practice, with the discovery of one new globular cluster, VVV CL001, reported on in \citet{dmea11}, and an additional three cluster candidates studied by \citet{cmbea11}, at least one of which is very likely a previously unknown globular cluster. A $JHK_S$ color-composite image of VVV CL001 is shown in Figure~\ref{fig:vvvcl001}, where one can see its close proximity to the previously known globular UKS~1. 

\begin{figure}[t]
\centering
    \includegraphics[width=0.875\textwidth]{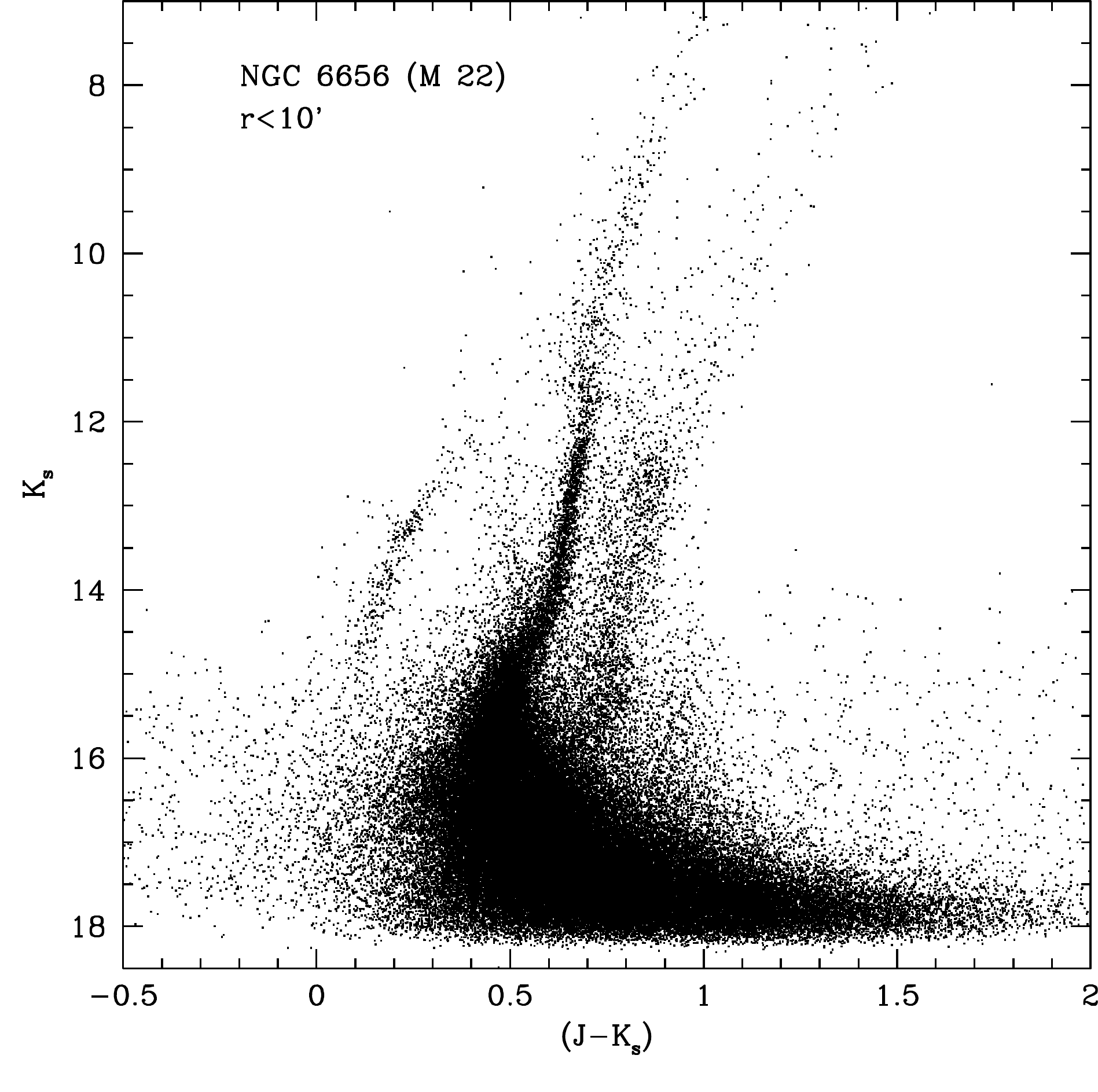}
\caption{CMD of the inner $10'$ region of NGC~6656 (M22). Note the prominent, seemingly non-uniformly populated blue horizontal branch. A prominent red giant branch and red horizontal branch associated with the Galactic bulge (in the background) can also be seen. 
\label{fig:m22}}
\end{figure}

\begin{figure}[t]
\centering
    \includegraphics[width=0.925\textwidth]{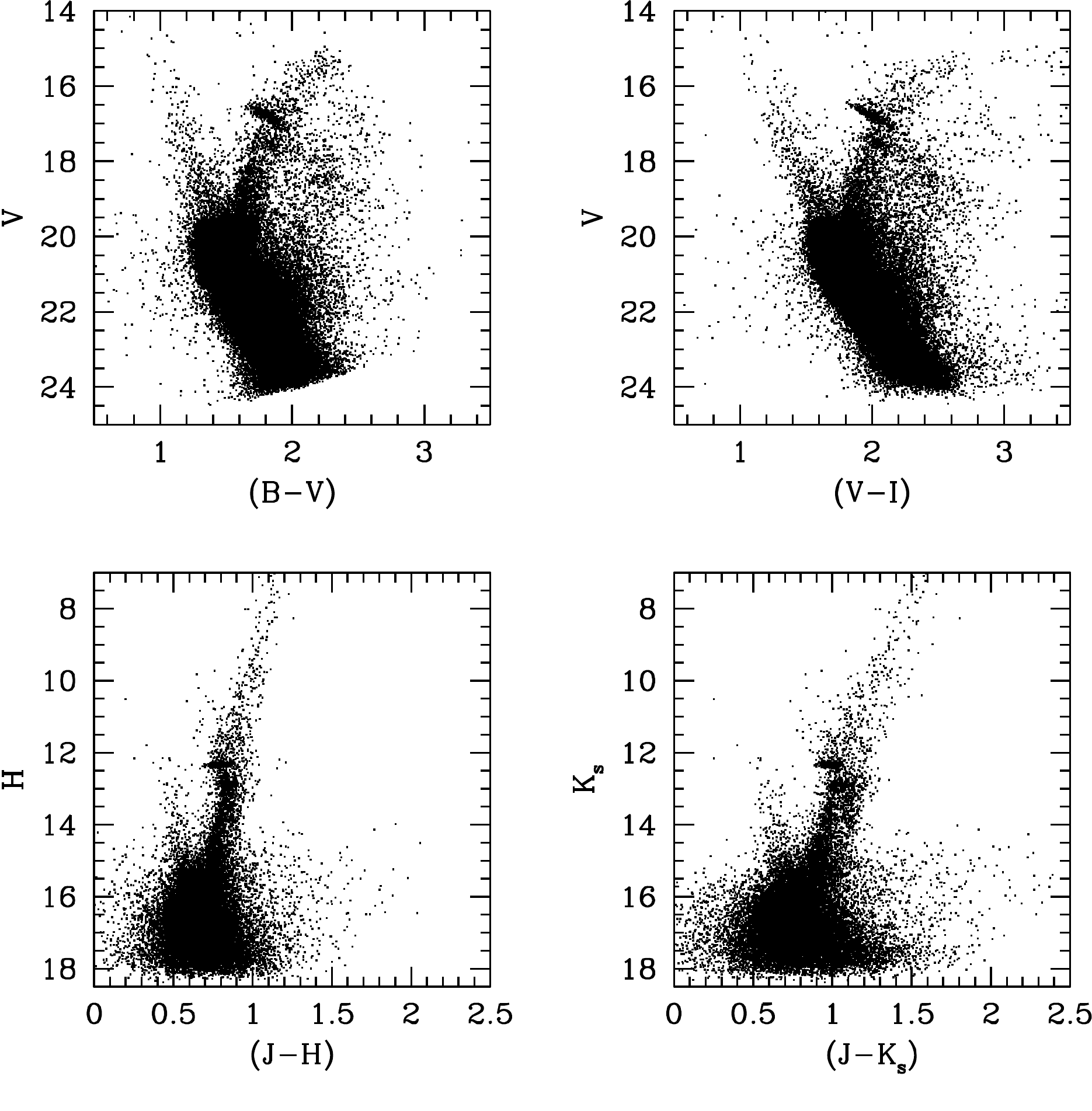}
\caption{Optical and near-IR CMDs of the inner $2.5'$ region of the globular cluster NGC~6553, obtained in the visual with ACS@HST and IMACS@Magellan ({\em top row}), and in the infrared with 2MASS and VVV data ({\em bottom row}). (2MASS data are used only to supplement the bright, saturated end of the VVV $JHK_S$ data~-- i.e., $K_S \lesssim 11$.) The effects caused by absolute and differential extinction are greatly reduced at infrared wavelengths, which allows a better comparison of the observational CMDs with theoretical models, thus leading to more accurate values of derived physical properties, such as photometric metallicity and age.
\label{fig:optical-nearir}}
\end{figure}

In addition to new globular clusters, the VVV Survey is also providing an unprecedented deep, wide-field view of the 37 previously known globular clusters located in its observed fields. To illustrate the quality of the derived near-IR CMDs, in Figure~\ref{fig:6440-6642} we show $K_S, \, (J-K_S)$ CMDs for the innermost $2'$ regions of a metal-rich (NGC~6440, with ${\rm [Fe/H] } = -0.36$) and a moderately metal-poor (NGC~6642, with ${\rm [Fe/H] } = -1.26$) globular cluster. In Figure~\ref{fig:m22} we show a similar CMD, for the massive, metal-poor globular cluster M22 (NGC~6656), with ${\rm [Fe/H] } = -1.70$, where a prominent blue horizontal branch extension is clearly observed, with a hint that it is far from being uniformly populated. Also, to illustrate the far superior handle on differential extinction that is provided by the VVV near-IR data, in Figure~\ref{fig:optical-nearir} we compare optical and near-IR CMDs of the globular cluster NGC~6553, with ${\rm [Fe/H] } = -0.18$. (Metallicities are from the Dec. 2010 version of the \citeauthor{wh96} \citeyear{wh96} catalog.) In fact, one of the main advantages of the VVV survey is that we can sample and study every cluster present in the surveyed region up to its tidal radius~-- and indeed beyond. This will allow us to more accurately define the structural parameters of these clusters, and to study their stellar population (or populations) out to regions not usually observed, due to the smaller fields of view of most instruments.

\begin{figure}[t]
\centering
    \includegraphics[width=\textwidth]{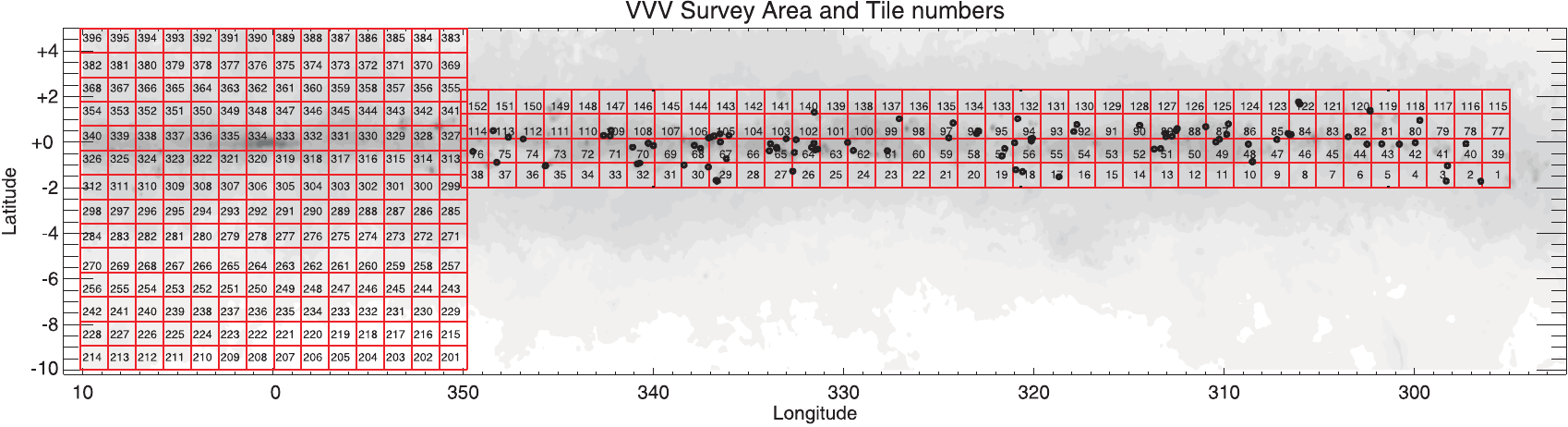}
\caption{As in Figure~\ref{fig:vvvfield}, but with the positions of the new star cluster candidates marked as dots. From \citet{jbea11}.
\label{fig:vvvarea}}
\end{figure}

\begin{figure}[t]
\centering
    \includegraphics[width=0.85\textwidth]{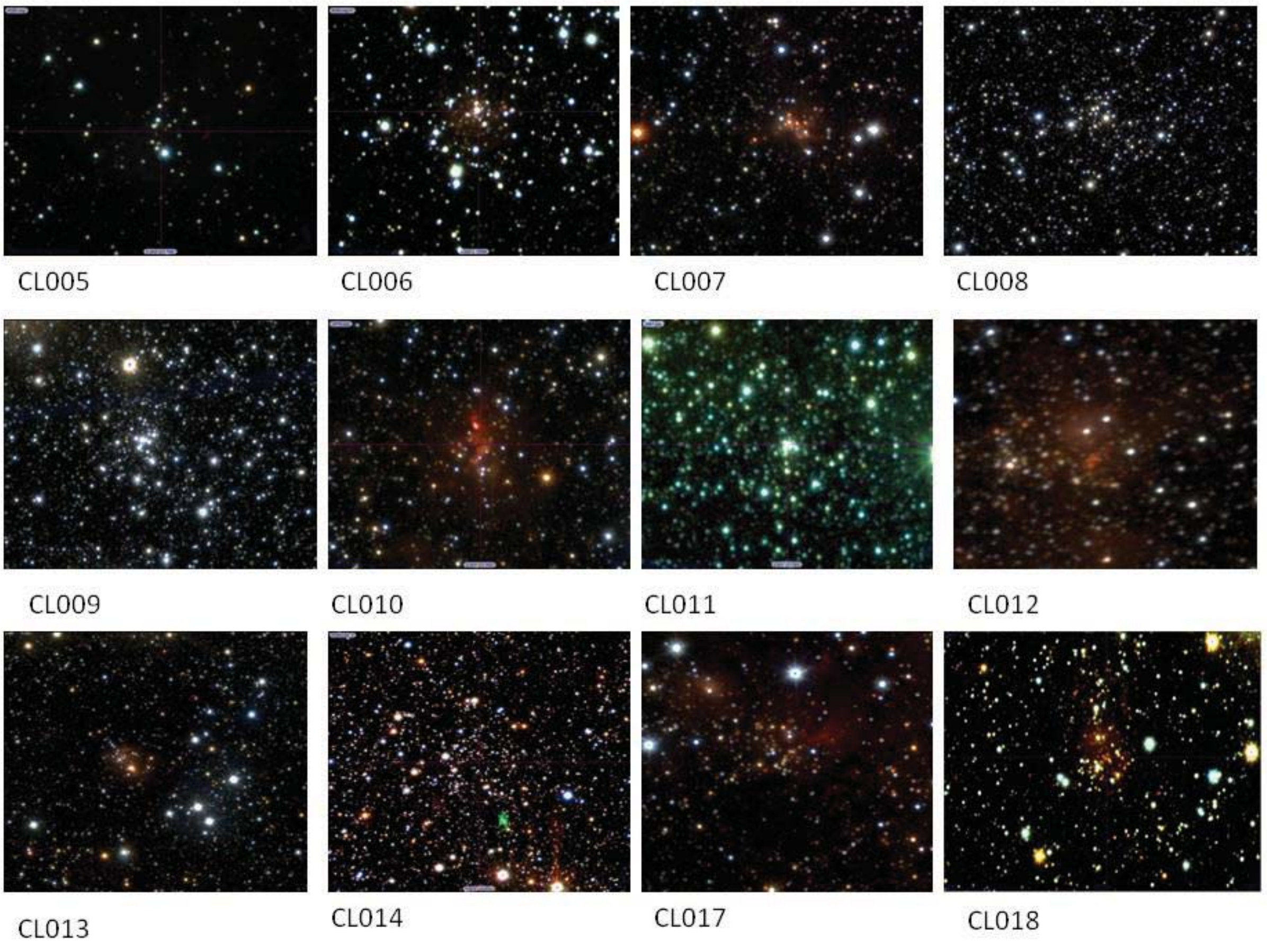}
    \includegraphics[width=0.85\textwidth]{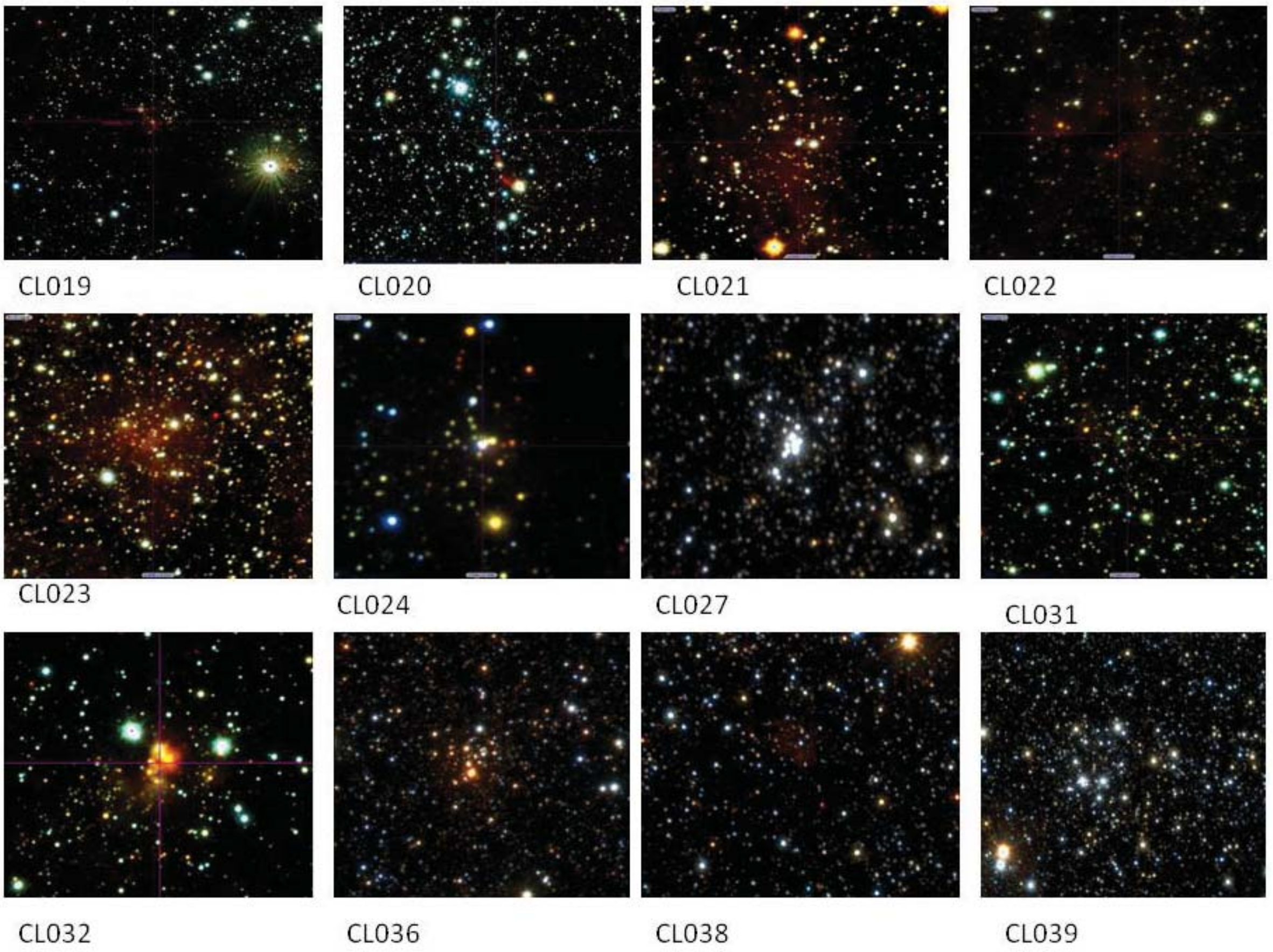}
\caption{$JHK_S$ true-color images of some of the new cluster candidates from the VVV Survey. The coordinates of the new candidates are given in \citet{jbea11}.
\label{fig:clusters}}
\end{figure}

\begin{figure}[ht]
\centering
    \includegraphics[width=\textwidth]{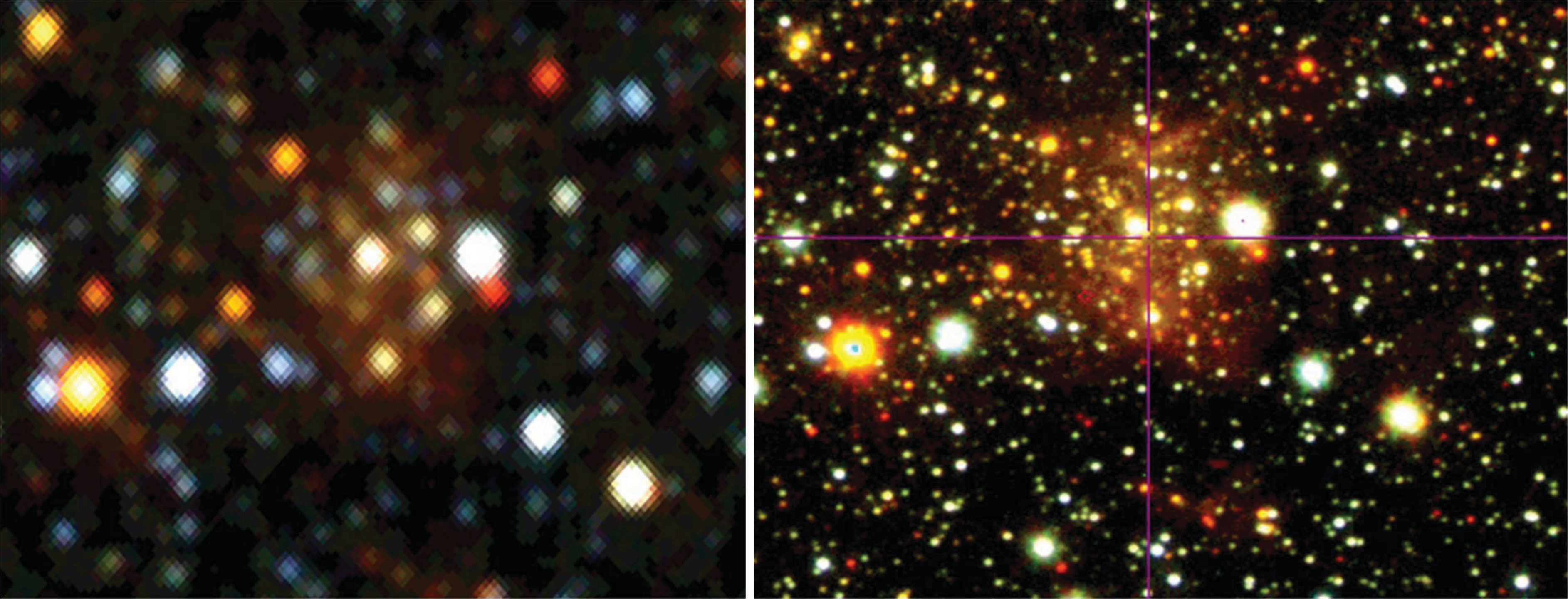}
\caption{2MASS ({\em left}) and VVV ({\em right}) composite $JHK_S$ color images of the cluster candidate VVV CL062 \citep{jbea11}. The field of view is approximately $2.2\arcmin \times 1.8\arcmin$. Note the far superior resolution of the VVV image.
\label{fig:vvv2mass}}
\end{figure}

\begin{figure}[h]
\centering
    \includegraphics[width=\textwidth]{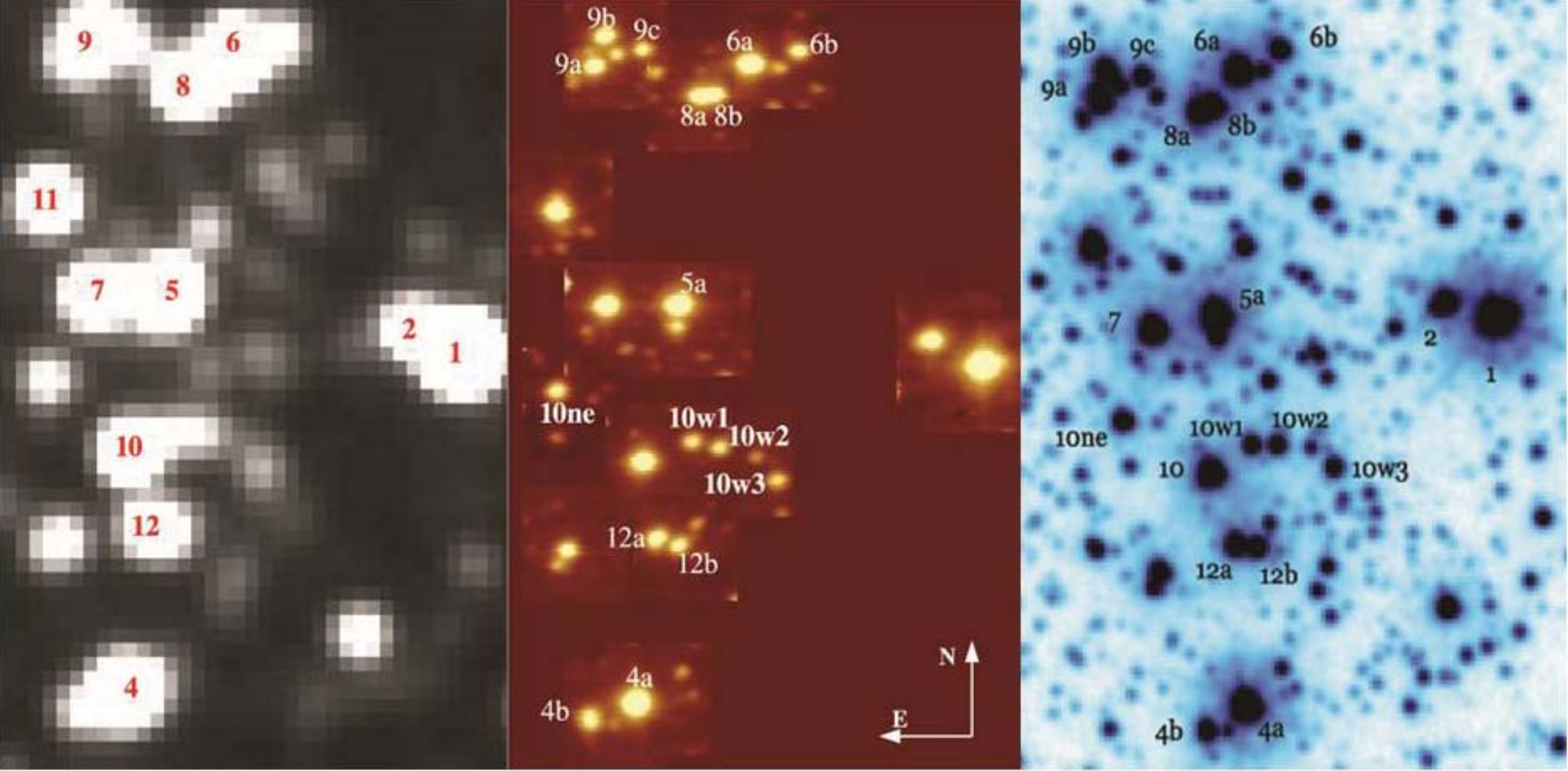}
\caption{2MASS ({\em left}), SINFONI@VLT ({\em middle}) and VVV ({\em right}) $K_S$-band images of the ionizing star cluster RCW79. The left and middle panels, as well as the identification of individual sources, are taken from \citet{fmea10}. Note that most of the sources resolved with SINFONI are also resolved in the VVV image.
\label{fig:vvvsinfoni}}
\end{figure}

\subsection{Open Clusters and Associations}\label{sec:open}

Estimates indicate that the Galaxy presently hosts 35,000 or more star clusters \citep{cbea06,pzea10}. However, only about 2500 open clusters have been identified, which constitute a sample affected by several well-known selection effects. Moreover, less than a half of these clusters have actually been studied, and this subset suffers from further selection biases. In this sense, the VVV Survey will not only lead to the discovery of a vast number of previously unknown open clusters and associations, but also~-- and importantly~-- allow systematic follow-up studies of their properties to be carried out. 

The first efforts by our team members to detect previously unknown open clusters and associations in the VVV Survey areas has led to the discovery of 96 candidates, which are reported on in \citet{jbea11}. The positions of these candidates are shown in Figure~\ref{fig:vvvarea}. Most of the new candidate clusters are faint and compact, are highly reddened, and likely younger than about 5~Myr. Composite color images of several of these candidates are shown in Figure~\ref{fig:clusters}. As can be seen from Figure~\ref{fig:vvv2mass}, where 2MASS and VVV images of the cluster candidate VVV CL062 are compared, detection of the new candidates has only become possible due to the much superior resolution of the VVV images, compared with 2MASS. Another such comparison is given in Figure~\ref{fig:vvvsinfoni}, where $K_S$-band images of the ionizing star cluster RCW79 taken with 2MASS (left panel), SINFONI@VLT (middle panel), and VVV (right panel) are compared. 

\begin{figure}[t]
\centering
    \includegraphics[width=\textwidth]{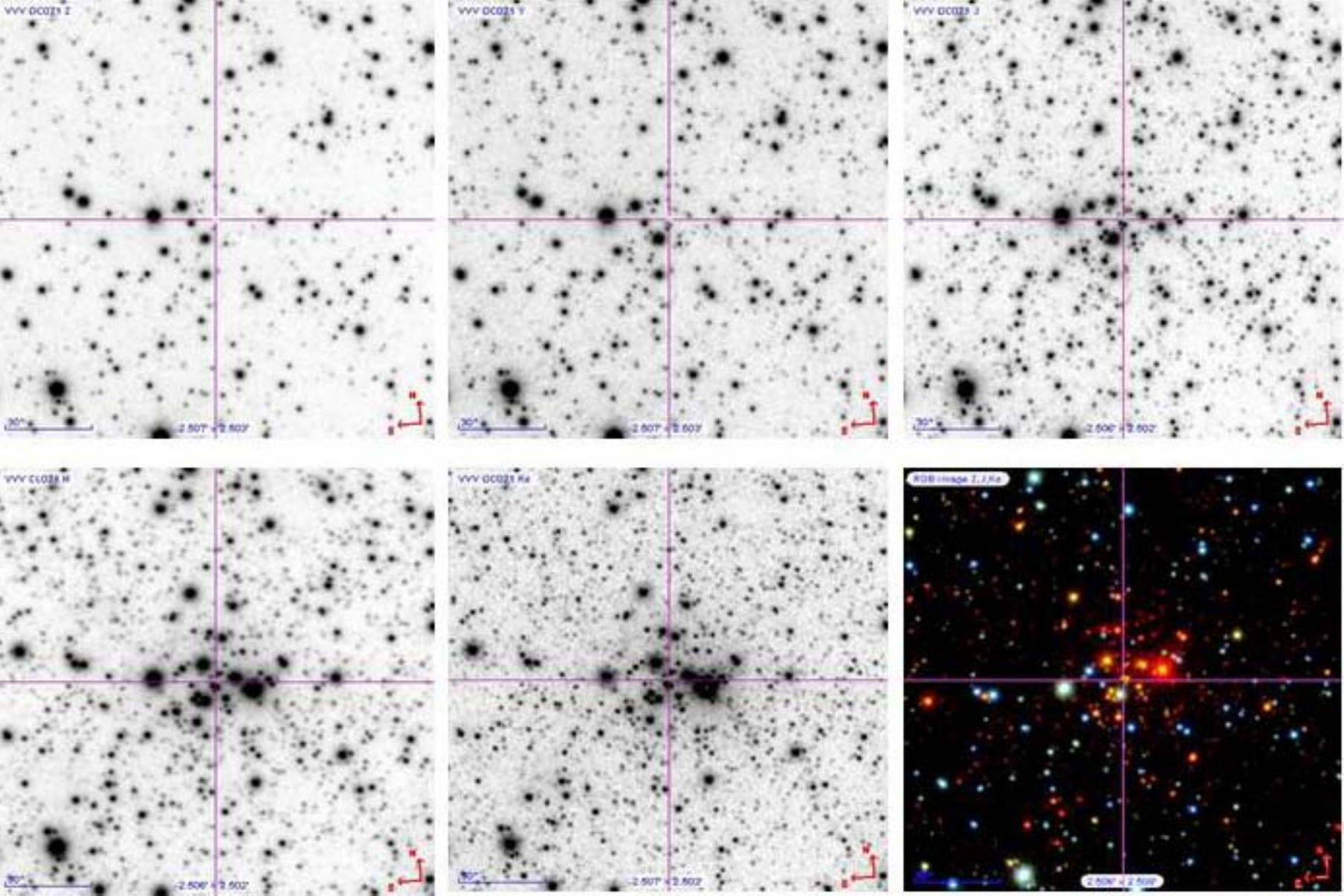}
\caption{VVV $ZYJHK_S$ images and $ZJK_S$ true-color image of VVV CL036. The field of view is $2.5' \times 2.5'$. North is up and East is to the left. From \citet{jbea11}.
\label{fig:ZYJHK}}
\end{figure}

\begin{figure}[t]
\centering
    \includegraphics[width=\textwidth]{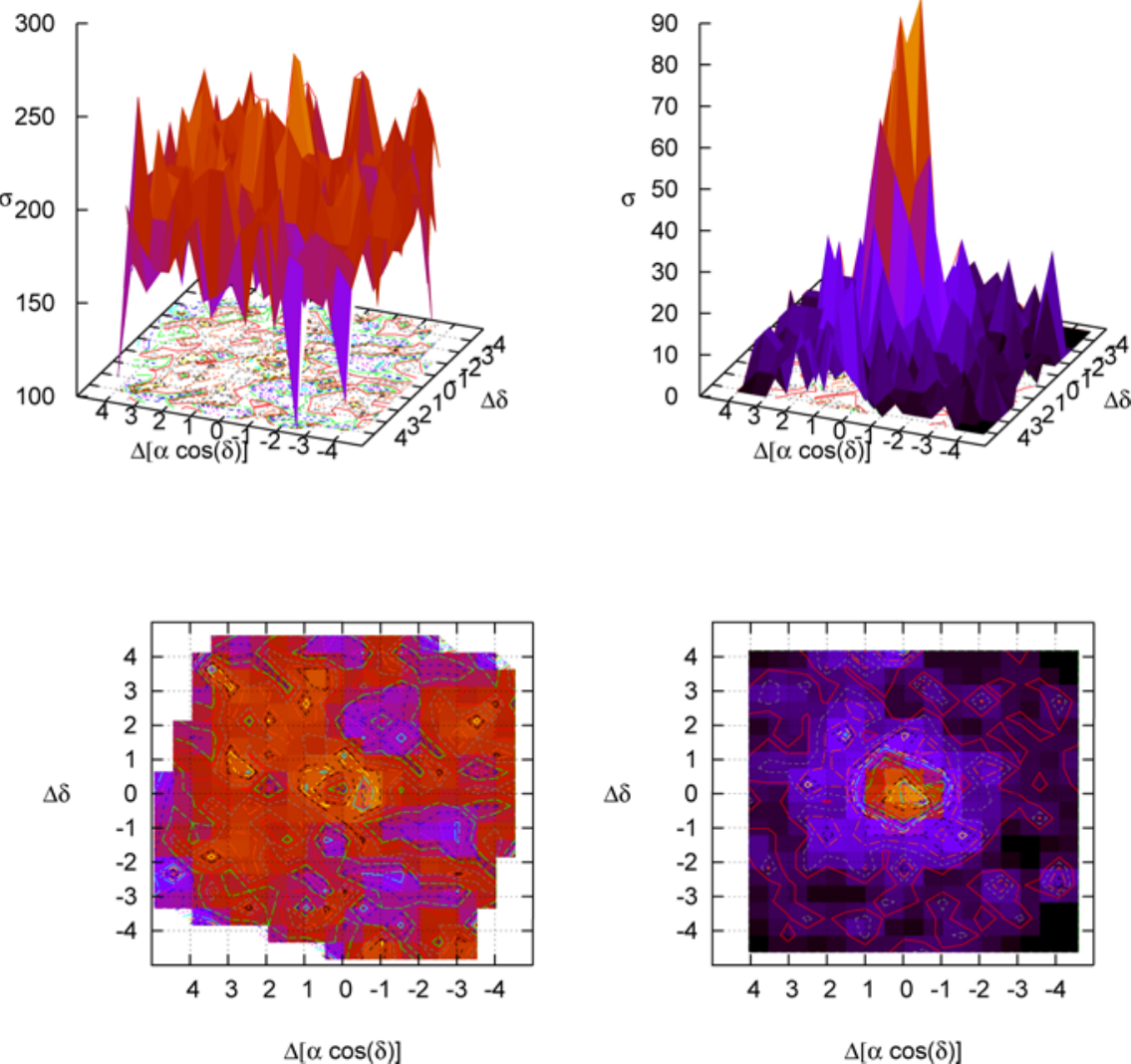}
\caption{Stellar surface density $\sigma$ (stars ${\rm arcmin}^{-2}$) map for VVV CL036.
On the left is shown the radial density profile produced with the raw photometry,  
whereas on the right the same field is shown after statistical decontamination. 
From \citet{jbea11}.
\label{fig:sigmaopen}}
\end{figure}

For one of our new cluster candidates, VVV CL036, a comparison between images taken in the $ZYJHK_S$ filters is shown in Figure~\ref{fig:ZYJHK}, clearly showing that the cluster only stands out in the near-IR. Even so, proper analysis of each cluster's CMD branches and derivation of the physical parameters requires careful statistical decontamination from the presence of field stars. In Figure~\ref{fig:sigmaopen} we illustrate the procedure adopted in \citet{jbea11} for the field star decontamination, which leads to excellent results, with the actual cluster population clearly standing out from the foreground/background field.

\subsection{Reddening-Free Indices}\label{sec:indices}

\begin{figure}[t]
\centering
    \includegraphics[width=0.875\textwidth]{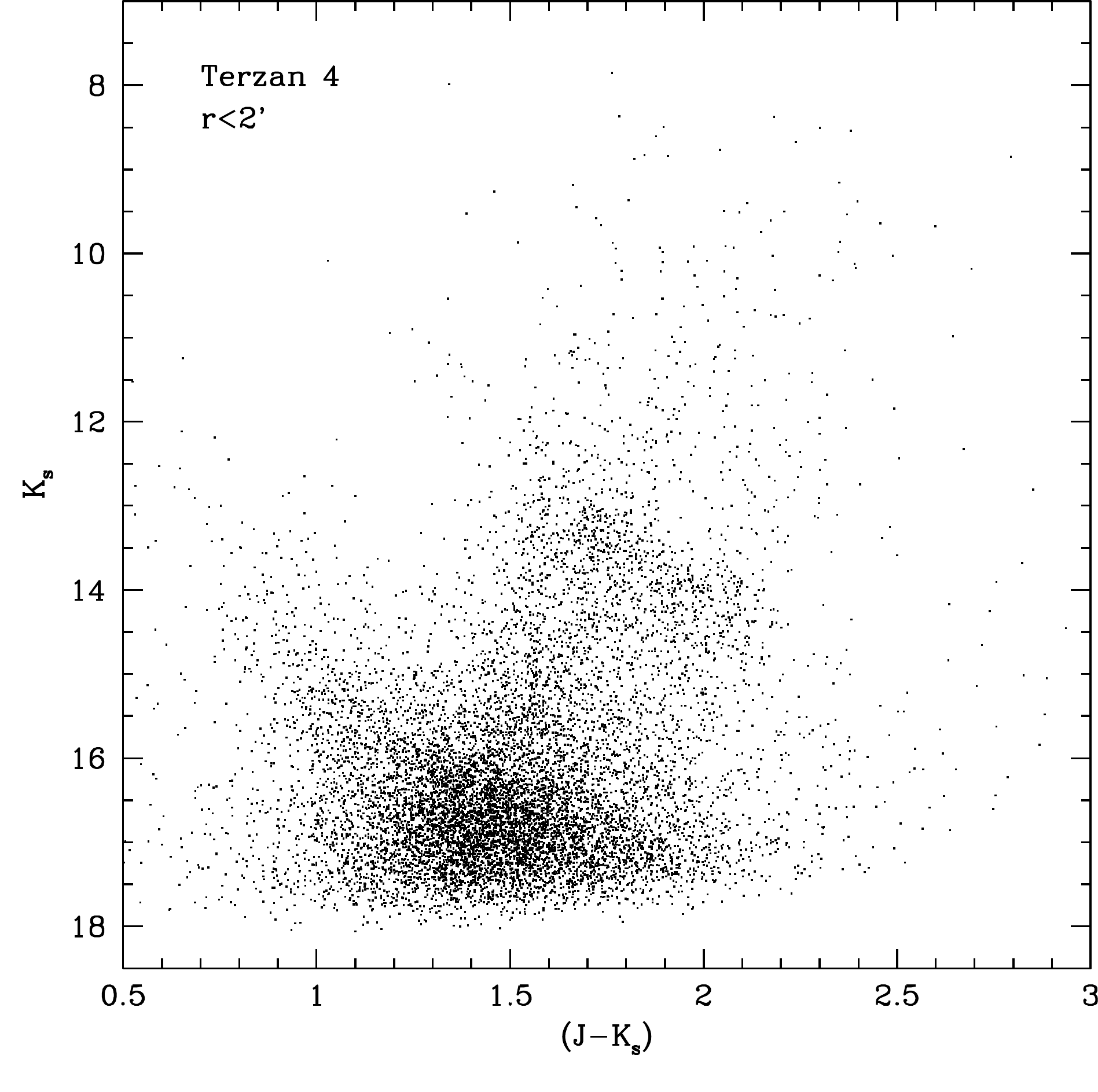}
\caption{CMD of the inner $2'$ region of Terzan 4. Even though the effects of the interstellar extinction are much reduced in the IR when compared with the optical, the high amounts  of dust and gas in the direction of the inner Galaxy can still affect the quality of the CMDs of some of the surveyed globular clusters. We plan to use techniques to decontaminate CMDs such as this one from differential extinction effects.
\label{fig:ter4}}
\end{figure}

Even in the near-IR filters used in the VVV Survey, differential reddening effects can often complicate the analysis of cluster CMDs. As an example, in Figure~\ref{fig:ter4} we show the near-IR CMD that we have derived for the moderately metal-poor (${\rm [Fe/H]} = -1.41$) globular cluster Terzan~4, whose reddening is, according to the \citet{wh96} catalog (Dec. 2010 version) as high as $E(\bv) = 2.0$. Clearly, in spite of using near-IR bandpasses, the cluster sequences are still severely distorted by differential reddening. 

The fact that the VVV Survey utilizes a set of 5 different broadband filters allows
us to explore the possibility of devising reddening-free indices, which can be very 
important for the studies of regions of high extinction (as is normally the case 
for the bulge and inner disk regions probed by this survey). To obtain such a set 
of reddening-free indices, we proceed as follows. 

First, we assume a standard extinction law, with $R = 3.09$ \citep{rl85}. We then adopt the effective
wavelengths $\lambda_{\rm eff}$ for the $ZYJHK_S$ filters that are listed in Table~\ref{tab:ext}. 
Based on these effective wavelengths, and using the \citet{jcea89} extinction law,\footnote{ 
See {\tt http://wwwmacho.mcmaster.ca/JAVA/Acurve.html}.} 
we obtain the relative extinctions for these filters that are listed in Table~\ref{tab:ext}.

These immediately lead to the following, reddening-free indices (among others): 

\begin{equation}
m_1 \equiv K_S - 1.08 \, (Z-Y),
\label{eq:m1}
\end{equation}

\begin{equation}
m_2 \equiv H - 1.13 \, (J-K_S),
\label{eq:m2}
\end{equation}

\begin{equation}
m_3 \equiv J - 1.03 \, (Y-K_S),
\label{eq:m3}
\end{equation}

\begin{equation}
m_4 \equiv K_S - 1.22 \, (J-H),
\label{eq:m4}
\end{equation}

\noindent which can all be used as ``pseudo-magnitudes,'' as reddening-free replacements for $J$, $H$, and $K_S$; and 

\begin{equation}
c_1 \equiv (Y-J) - 1.14 \, (J-H),
\label{eq:c1}
\end{equation}

\begin{equation}
c_2 \equiv (Z-Y) - 0.99 \, (Y-J),
\label{eq:c2}
\end{equation}

\begin{equation}
c_3 \equiv (J-H) - 1.47 \, (H-K_S),
\label{eq:c3}
\end{equation}

\begin{equation}
c_4 \equiv (J-K_S) - 1.50 \, (Z-Y),
\label{eq:c4}
\end{equation}

\noindent which can all be used as ``pseudo-colors,'' or reddening-free replacements for such colors as $(Y-J)$, $(Z-Y)$, $(J-H)$, $(H-K_S)$, and $(J-K_S)$.

\begin{flushleft}
\begin{deluxetable*}{rccc}
\tabletypesize{\normalsize}
\tablecaption{Relative Extinctions for $ZYJHK_S$ Filters (VISTA Set)}
\tablewidth{0pt}
\tablehead{ \\ \colhead{Filter} & \colhead{$\lambda_{\rm eff}$ ($\mu$m)}  & \colhead{$A_X/A_V$} &  \colhead{$A_X/E(\bv)$} \\
}
\startdata
$Z$   & 0.878 & $0.499$ & $1.542$ \\ 
$Y$   & 1.021 & $0.390$ & $1.206$ \\ 
$J$   & 1.254 & $0.280$ & $0.866$ \\ 
$H$   & 1.646 & $0.184$ & $0.567$ \\ 
$K_S$ & 2.149 & $0.118$ & $0.364$ \\
\enddata
\label{tab:ext}
\end{deluxetable*}
\end{flushleft}

As a note of caution, the reader should also bear in mind that, when using such indices, the errors propagate accordingly. In addition, the intrinsic range in values (in mag units) covered by different CMD features can be significantly reduced, with respect to that seen in the original CMDs. Therefore, in order for clear-cut sequences to show up in such reddening-free diagrams, the internal precision must be significantly higher than what is needed for similar such sequences to be seen in the original CMDs. Note also that errors in the calibrations, particularly in the color terms, may also contribute to smearing up features that might otherwise show up well in such reddening-free diagrams. Last but not least, it is important to bear in mind that eqs.~(\ref{eq:m1}) through (\ref{eq:c3}) were obtained under the assumption of a standard extinction law with $R = 3.09$. For applications to regions with different $R$ values, these reddening-free indices should be redefined accordingly.

\begin{figure}[t]
\centering
    \includegraphics[width=0.925\textwidth]{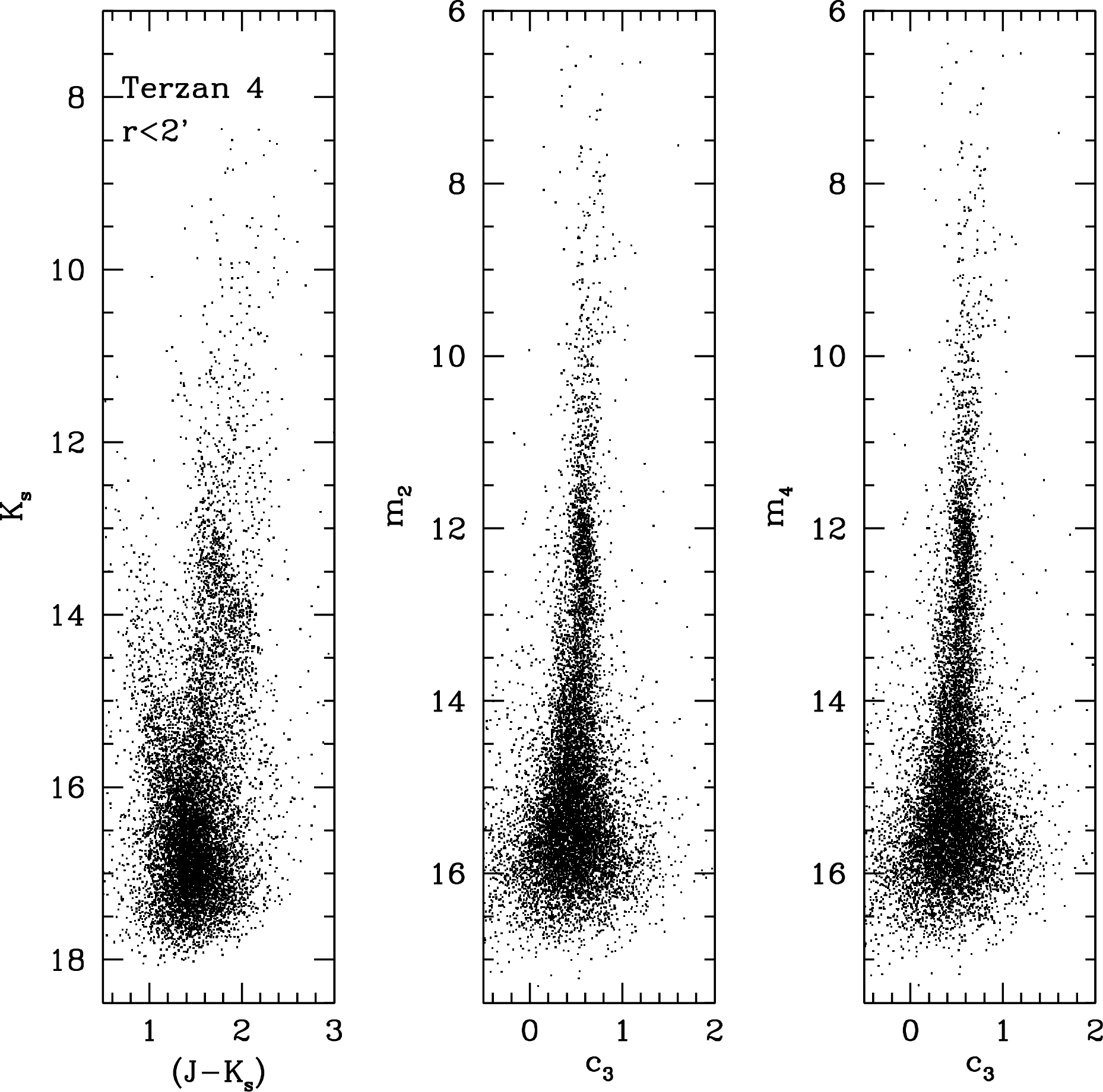}
\caption{Same as in Figure~\ref{fig:ter4}, but using our newly-defined, reddening-free indices $m_2$ (eq.~\ref{eq:m2}), $m_4$ (eq.~\ref{eq:m4}), and $c_3$ (eq.~\ref{eq:c3}). The {\em left panel} shows the original $K_S$, $(J-K_S)$ CMD, whereas the {\em middle panel} shows the $m_2, \, c_3$ CMD, and the {\em right panel} shows the $m_4, \, c_3$ CMD. Note the remarkable tightening of the cluster's sequences, when the reddening-free indices are used. 
\label{fig:ter4reddfree}}
\end{figure}

\section{The VVV Survey Templates Project}\label{sec:templates}

As already pointed out (\S\ref{sec:intro}), the VVV ESO Public Survey will produce a catalogue and light curves for an estimated $\sim 10^6$ variable stars, most of which will be previously unknown. Obviously, this very large number of light curves cannot be classified using traditional methods, particularly visual inspection. Instead, an automated classification scheme must be developed in order to properly deal with the classification of these variable stars. This is the ultimate goal of the VVV Templates Project.\footnote{\tt{http://www.vvvtemplates.org}}

\subsection{The Need for Near-IR Light Curve Templates}\label{sec:needtemplates}

Unlike previous variability surveys, the VVV Survey is carried out in the near-IR. Despite several fundamental advantages, mostly due to the ability to probe deeper into the heavily reddened regions of the Galactic bulge and plane, the use of this spectral region also presents us with important challenges. In particular, the high-quality templates that are needed for training the automated variable star classification algorithms are not available. This is very different from the situation of variability surveys carried out in the visible, for which vast quantities of high-quality template light curves are available \citep[see, e.g.][]{jdea07,jblea11,pdea11}. 

To properly illustrate the importance of this point, one should consider that not only is the currently available number of near-IR templates inadequate for our purposes (since they do not properly sample the large variety of light curve shapes encountered for many of the most important variability classes). Also, many such classes have not yet been observed in a sufficiently extensive way in the near-IR, so that good light curves are entirely lacking for these classes. Therefore, in order to properly classify the $\sim 10^6$ light curves produced by the VVV Survey, we need to build our own template light curves.

\begin{figure}[t]
\centering
    \includegraphics[width=0.95\textwidth]{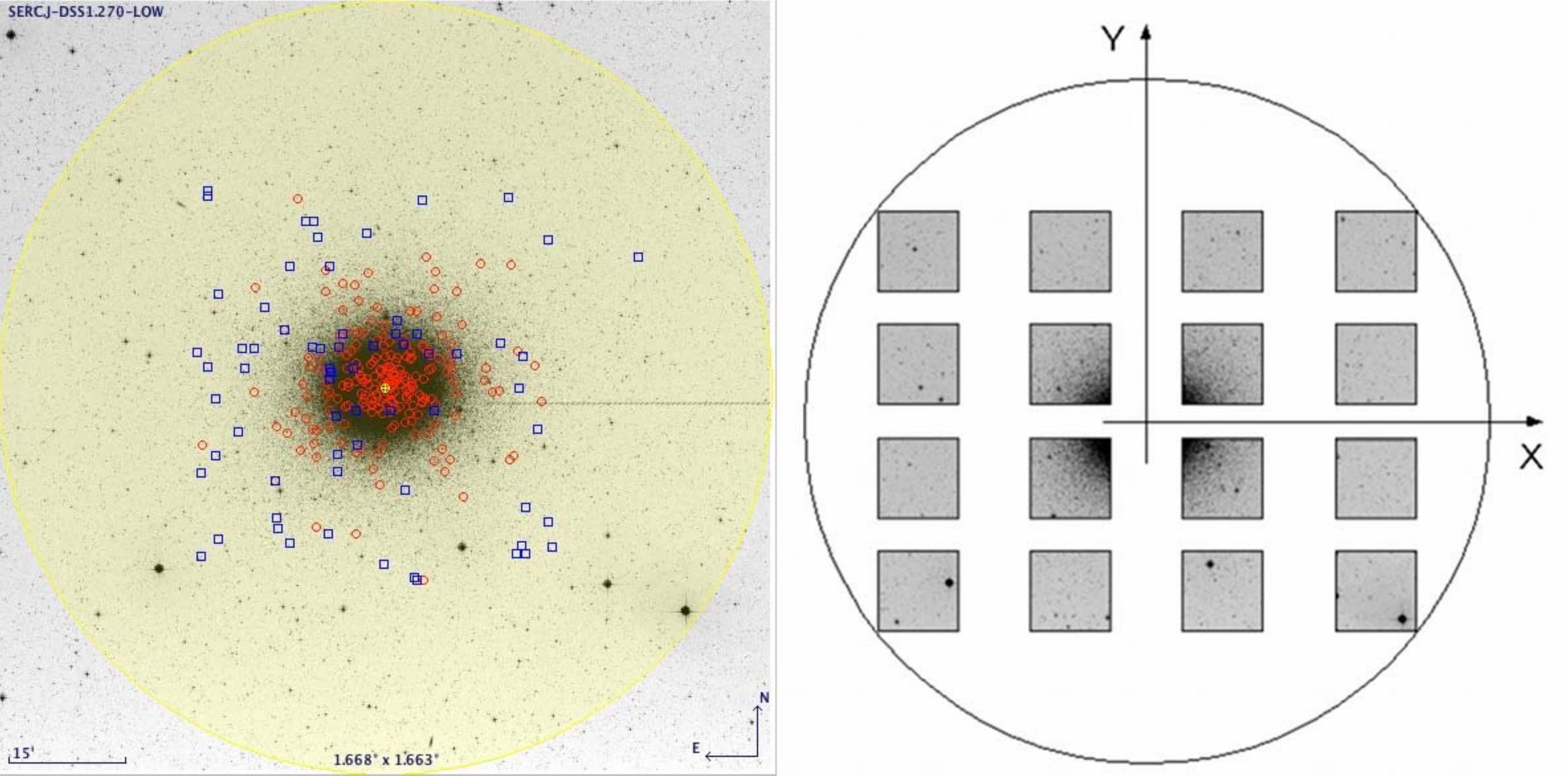}
\caption{Upcoming VISTA observations of the globular cluster $\omega$~Cen. {\em Left panel}: An example of variable star distribution across the cluster field: red circles mark the positions of known RR Lyrae stars, while blue squares mark the positions of known eclipsing binaries. {\em Right panel}: VIRCAM@VISTA's 16 detectors, with $\omega$~Cen at the center of the focal plane.
\label{fig:omegacen}}
\end{figure}

\subsection{Current Status and Future Perspectives}\label{sec:statustemplates}

In the more general framework of the VVV Survey, the VVV Templates Project has turned out to be a large observational effort in its own right, aimed at creating the first database on stellar variability in the near-IR, i.e. producing a large database of well-defined, high-quality, near-IR light curves for variable stars belonging to different variability classes. By monitoring hundreds of (optically well-studied) variable stars in the $JHK_S$ bands, its primary goal is thus to provide a statistically significant training set for the automated classification of VVV light curves (see \S\ref{sec:automated}).

The astronomical community as a whole has been very supportive, with the end result that we have secured time for this project using several IR facilities across the globe, as listed in Table~\ref{tab:templates}. The availability of telescope time is key to ensure that we produce a significant number of template-grade light curves in time for classification of the VVV light curves, i.e. within the next 3 years or so, before the bulk of the VVV variability campaign is carried out \citep{dmea10,rsea10}. In order to ensure a homogeneous observational strategy and to optimize the use of the awarded time, each telescope/instrument combination is used to build template light curves for at most a very few specific variability classes, as can also be seen from Table~\ref{tab:templates}. 

Importantly, the scientific return of the VVV template light curves will not be restricted to the automated classification of the $\sim 10^6$ light curves produced by the VVV Survey. Rather, these light curves will provide us with a unique opportunity to expand our knowledge of the stellar variability phenomenon per se, by tackling the comparatively ill-explored near-IR regime. 

Consider, as an example, our upcoming monitoring, using the VISTA telescope, of the variable star content of $\omega$~Centauri (Fig.~\ref{fig:omegacen}). Such a project will lead to high-quality light curves for more than 250 variables, greatly exceeding, both in quantity and in completeness of each individual light curve, what was achieved in previous near-IR studies of the cluster \citep{dpea06}. In particular, since $\omega$~Cen hosts the largest known population of SX~Phoenicis stars, some 75 members, we will be in a position to directly derive, for the first time, a period-luminosity relation for SX~Phe stars in the near-IR \citep{fs03,dwea07}. In like vein, we will also improve the calibration of the RR Lyrae period-luminosity relation in the near-IR, given that we will obtain, for the first time, complete light curves for most of the 170 RR Lyrae variables in the cluster.

Similarly, we have recently started to monitor a series of open clusters known to host sizeable populations of $\delta$~Scuti stars. In addition to high-quality near-IR templates, this will also allow us to verify the period-luminosity relation in the near-IR that has been suggested for this important class of variables \citep{jk90}.

\begin{figure}[t]
\centering
    \includegraphics[width=0.975\textwidth]{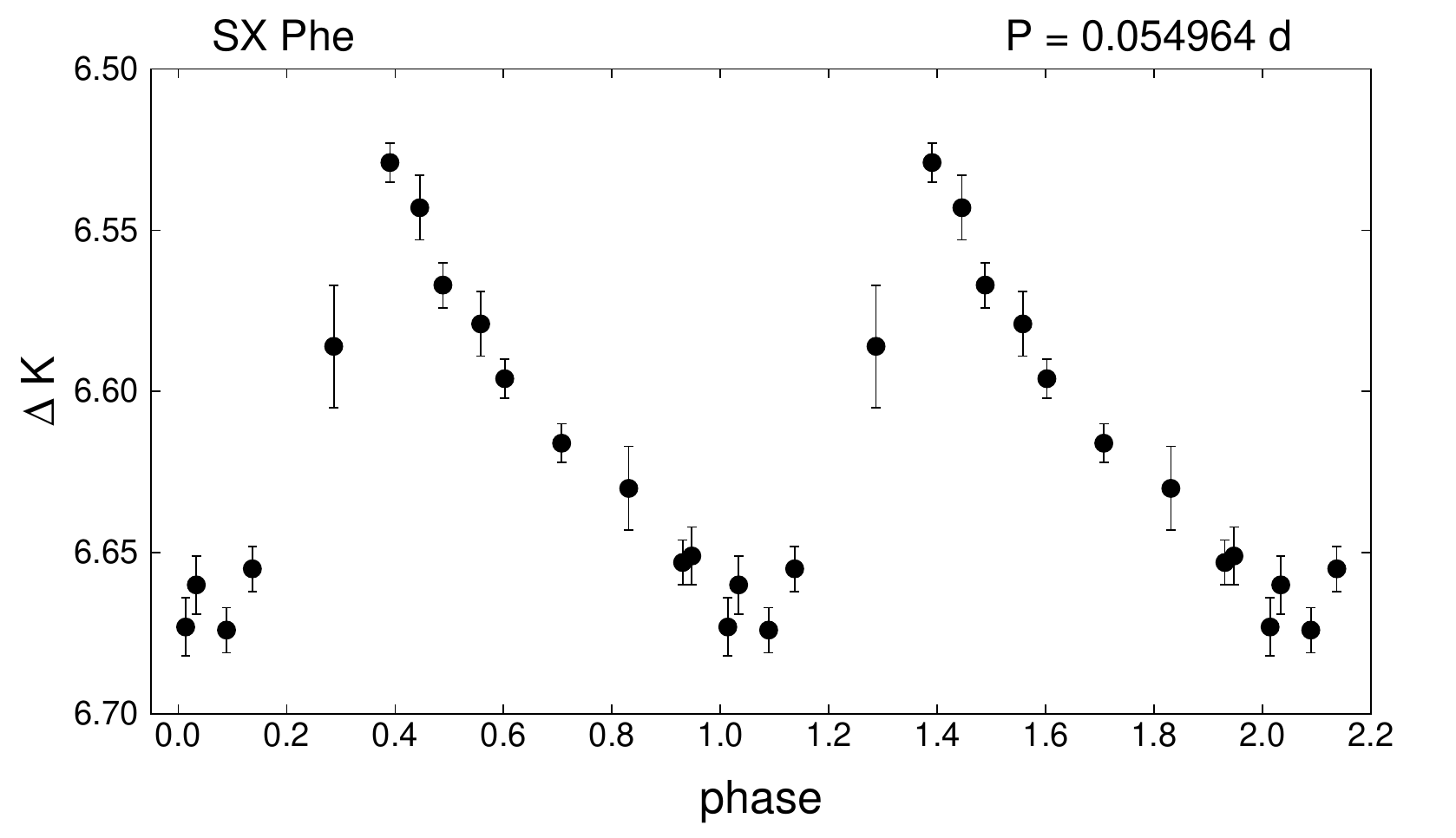}
\caption{$K$-band light curve of SX~Phe, taken at the SAAO 0.75m telescope in Sept. 2010. Note the overall amplitude of $\approx 0.15$~mag. 
\label{fig:sxphe}}
\end{figure}

\begin{figure}[t]
\centering
    \includegraphics[width=0.975\textwidth]{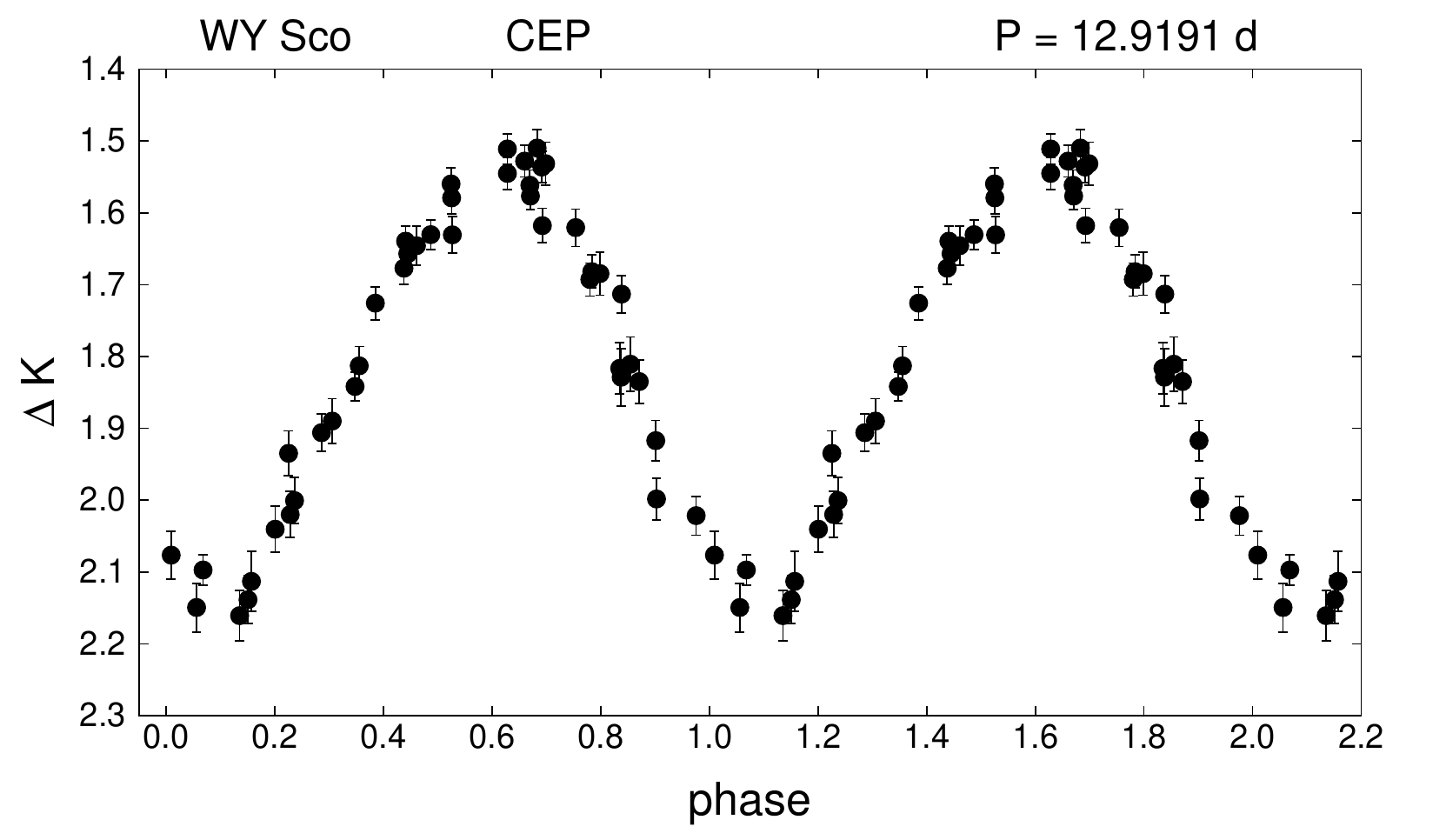}
\caption{$K$-band light curve of the classical Cepheid WY~Sco, currently being observed with the REM 0.6m telescope in the framework of the VVV Templates Project. 
\label{fig:rem}}
\end{figure}

Some of the first light curves obtained in the framework of the VVV Templates Project are shown in Figure~\ref{fig:sxphe} ($K$-band light curve for SX~Phe, the prototype of its class) and Figure~\ref{fig:rem} ($K$-band light curve for WY~Sco, a classical Cepheid). The SX~Phe light curve was obtained in the course of an observing campaign carried out with the SAAO 0.75m photometer in Swinburne, South Africa, whereas the WY~Sco light curve has been obtained in the course of our ongoing observations with the REM 0.6m robotic telescope in La Silla, Chile. By the end of our monitoring campaigns for these and other stars, the number of data points per light curve will at least double, as required in order to properly define light curve templates.

\begin{flushleft}
\begin{deluxetable*}{rccccc}
\tabletypesize{\normalsize}
\tablecaption{VVV Templates Project: Observing Time at Different Observatories}
\tablewidth{0pt}
\tablehead{ \\ \colhead{Observatory}   & \colhead{Telescope} &  \colhead{IR Camera}  & \colhead{FoV}  & \colhead{Awarded Time}  & \colhead{Target} \\
}
\startdata
ESO Paranal & VISTA 4m & VIRCAM & $60'\times90'$ & 34 hours & $\omega$ Centauri\\
CTIO & Blanco 4m & NEWFIRM & $28'\times28'$ & 3 nights & NGC 3293/4755/6231\\
KASI & BOAO 1.8m & KASINICS & $3.6'\times3.6'$ & 6 weeks & CVs, $\delta$ Scutis \\
IAC & TCS 1.4m & CAIN-III & $4.2'\times4.2$ & 36 nigths & NGC 1817/7062 \\
SAAO & IRSF 1.4m & SIRIUS & $7.7'\times7.7'$ & 3 weeks & M62, NGC 1851/6134\\
CTIO & SMARTS 1.3m & ANDICAM & $2.4'\times2.4'$ & 90h/sem & field RR Lyrae\\ 
SAAO & 0.75m & MkII phot & -- & 3 weeks & SX Phe, Ellipsoidal, ...\\
ESO La Silla & REM & REMIR & $10'\times10'$ & 225h/sem & EBs, $\delta$ Scutis\\
Asiago${\tablenotemark{a}}$ & Schmidt & Opt. CCD & $52'\times36'$ & 36 nights & NGC 1817/7062\\
Kazakshtan${\tablenotemark{b}}$ & 1m  & Opt. CCD & $30'\times30'$ & 6 weeks & CVs\\
OMM${\tablenotemark{c}}$ & 1.6m  & CAPAPIR & $30'\times30'$ & 8 hours & NGC 7062\\
OAGH${\tablenotemark{c}}$ & 2.1m  & CANANEA & $4'\times4'$ & 10 nights & NGC 1817 \\
\enddata
\tablenotetext{a}{Simultaneous optical monitoring of TCS targets.}
\tablenotetext{b}{Simultaneous optical monitoring of BOAO targets.}
\tablenotetext{c}{Proposals submitted but still under evaluation by the respective TACs.}
\label{tab:templates}
\end{deluxetable*}
\end{flushleft}

\subsection{Towards the Automated Classification of VVV Light Curves}\label{sec:automated}

In the last few years, there has been an increasing interest towards the application of artificial intelligence algorithms in astronomical research. This has been mainly triggered by the ongoing and future large-scale observational surveys, which are expected to deliver huge amounts of high-quality data, which will need to be promptly analyzed and classified.  

An illustrative example of such a need is indeed the VVV Survey, whose final product will be a deep near-IR atlas in five passbands and a catalog with an estimated more than $10^6$ variable point sources \citep{dmea10,rsea10}. 

Since the majority of these surveys will deal with temporal series, it is natural that most of the effort for developing, adapting and testing such automated classification algorithms has concentrated on the field of variable stars. The reference work in machine learning applied to variable star classification can be considered the paper by \citet{jdea07}, in which the authors explored several classification techniques, quantitatively comparing performance (e.g., in terms of computational time) and final results (e.g., in terms of accuracy). More recently, a few other studies have focused on specific methodologies, with the implicit goal of finding the best compromise between robustness and speed \citep[e.g.,][]{jblea11,pdea11,jrea11}. 

Irrespective of the specific algorithm, the general idea behind these ``supervised machine learning'' methodologies is to create a function, \textit{the classifier}, that is able to infer the most probable \textit{label} for an object~-- in our case, the variability class to which an unclassified variable star belongs to~-- on the basis of what is learned by the analysis of the \textit{inputs}~-- in our case, light curve features, including, for instance, periods, Fourier decomposition parameters, colors, etc.~-- from a \textit{training set}. The latter, in our case, is a collection of high-quality light curves of independently classified variable stars~-- the {\em templates}.

\begin{figure}[t]
\centering
    \includegraphics[width=0.75\textwidth]{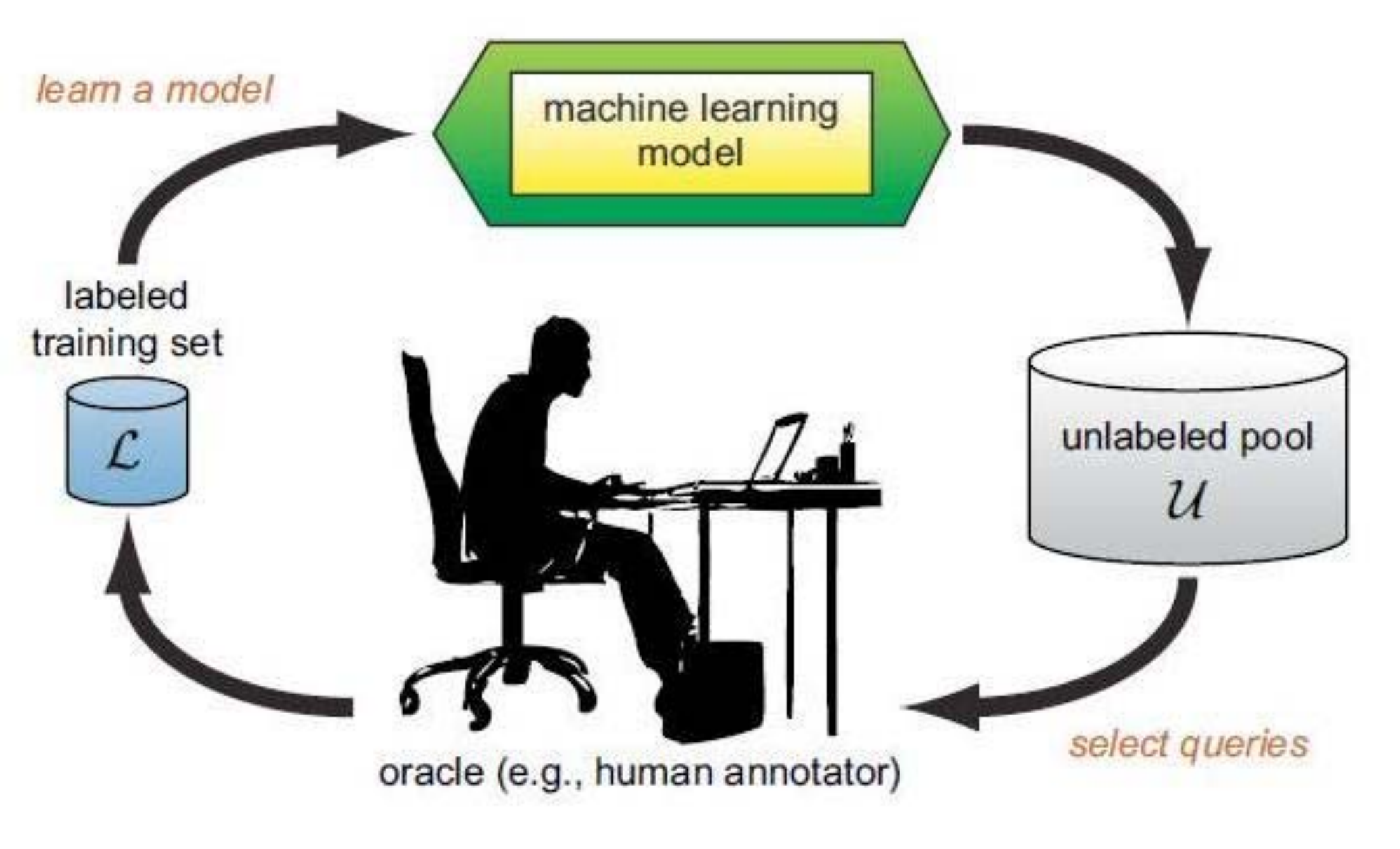}
\caption{The pool-based active learning cycle \citep[adapted from][]{bs10}.
\label{fig:altemplates} }
\end{figure}

In this sense, in \citet{kpea11} we show that a significant improvement in the automated classification of astronomical data can be achieved from an optimization of the way {\em the training set itself} is selected. This is of crucial importance for our purposes, given that the many hundreds of light curve templates that optical automated classification studies can rely upon \citep[see, e.g., Table~7 in][]{jrea11} are not yet available in the near-IR, and will likely~-- despite our best efforts~-- not be available in comparable numbers in the near future. Given the availability of such huge template datasets, the precise way in which the training set itself is optimized has never been a major concern in optical studies, but is instead an important aspect of our near-IR automated classification work that must be carefully addressed. 

In fact, there are concrete cases in which it is convenient to keep the training set as small as possible, while still retaining its maximum information content. For example, one may need to classify {\em as quickly as possible} huge amounts of data obtained in ``real-time,'' such as the 30~TB of data produced every night by LSST \citep{ziea08}. We might also be in the situation for which gathering a training set may be far from straightforward, as in the case of the VVV Survey. In all these cases, an active learning approach can be the key to the problem. 

Active learning (Fig.~\ref{fig:altemplates}) is a form of supervised machine learning in which the learning algorithm is able to interactively query the user (or some other information source). The key hypothesis is that if the learning algorithm is allowed to choose the data from which it learns, it will perform better with less training. In other words, the active learner aims to achieve high accuracy using as few (in our case) template light curves as possible, thereby minimizing the cost of obtaining observational data. Active learning is thus fully justified in many modern machine learning problems where unlabeled data may be abundant (in our case, e.g., the $10^9$ point sources in the VVV Survey catalog), but labels (i.e., template light curves) are scarce or expensive to obtain. This is the technique that we eventually plan to implement, in order to properly carry out the  automated classification of VVV light curves \citep{kpea11}.

\section{Summary}\label{sec:conclusions}
The VVV Survey of the Galactic bulge and inner disk, which has recently started its second year of observations, represents a treasure trove for many different types of studies. Its final data products will include a deep, multi-band catalog containing of order $10^9$ point sources in the $ZYJHK_S$ filters, in addition to near-IR light curves for an estimated $10^6$ variable stars in the surveyed fields. In this contribution, we have provided an overview of the project's main scientific goals and present status. We have also highlighted some of the first scientific results that have been obtained on the basis of its first year of operation, including the discovery of a number of previously unknown globular clusters, open clusters, and stellar associations. Deep color-magnitude diagrams for a few selected systems were presented, and a set of reddening-free indices for analysis of heavily differentially obscured regions introduced. $K_S$-band light curves for a few variable stars in the VISTA science verification field were also presented. Finally, we provided an overview of the VVV Templates Project, whose aim is to provide high-quality light curves for the automated classification of VVV near-IR light curves. In this framework, we also argued that active learning may represent an ideal technique for classifying the VVV light curves on the basis of the template set that we are currently building.

\vskip 1cm

\noindent {\bf Acknowledgments.} The VVV Survey is supported by the European Southern Observatory, 
the BASAL Center for Astrophysics and Associated Technologies (PFB-06), the FONDAP Center for 
Astrophysics (15010003), and the Chilean Ministry for the Economy, Development, and Tourism's 
Programa Iniciativa Cient\'{i}fica Milenio through grant P07-021-F, awarded to The Milky Way 
Millennium Nucleus. Support for M.C., I.D. and J.A.-G. is also provided by Proyecto Fondecyt 
Regular \#1110326. Support for R.A. is provided by Proyecto Fondecyt \#3100029. J.B. acknowledges
support from Proyecto Fondecyt Regular \#1080086. 


\end{document}